\documentclass[10pt]{amsart}

\usepackage[T1]{fontenc}

\usepackage{color}
\usepackage{amsmath}
\usepackage{amssymb}
\usepackage{float}
\usepackage{graphicx}

\newsavebox{\savepar}

\def\F{\mathcal{F}}

\def\S{\mathcal{S}}

\def\GG{{\frak G}}

\def\epsilon{\varepsilon}
\def\leq{\leqslant}

\def\x{{\bf x}}
\def\y{{\bf y}}

\def\u{{\bf u}}

\def\int{{\rm int}}

\def\GG{{\frak G}}

\def\u{{\bf u}}

\def\S{{\bf S}}
\def\F{{\bf F}}
\def\x{{\bf x}}
\def\y{{\bf y}}

\def\X{{\bf X}}

\def\BSigma{{\bf \Sigma}}

\def\Phat{\hat{P}}

\newcommand{\rem}[1]{}

\begin{document}
\title{From Microscopic Heterogeneity to Macroscopic Complexity in the Contrarian Voter Model}

\maketitle

\begin{center}
\begin{minipage}{13truecm}
%\begin{small}
\author{Sven Banisch\footnote{The author acknowledges financial support of the European Community's Seventh Framework Programme (FP7/2007-2013) under grant agreement no.~318723 (\emph{MatheMACS} -  {\tt http://www.mathemacs.eu}).}}

Max Planck Institute for Mathematics in the Sciences (Germany)\\
MPI-MIS, Inselstrasse 22, D-04103 Leipzig, Germany\\
eMail: sven.banisch@UniVerseCity.de

%\end{small}
\end{minipage}
\end{center}

\markboth{Banisch}{From Microscopic Heterogeneity to Macroscopic Complexity in the Contrarian Voter Model}

\begin{abstract}
An analytical treatment of a simple opinion model with contrarian behavior is presented. 
The focus is on the stationary dynamics of the model and in particular on the effect of inhomogeneities in the interaction topology on the stationary behavior.
We start from a micro-level Markov chain description of the model.
Markov chain aggregation is then used to derive a macro chain for the complete graph as well as a meso-level description for the two-community graph composed of two (weakly) coupled sub-communities.
In both cases, a detailed understanding of the model behavior is possible using Markov chain tools.
More importantly, however, this setting provides an analytical scenario to study the discrepancy between the homogeneous mixing case and the model on a slightly more complex topology.
We show that memory effects are introduced at the macro level when we aggregate over agent attributes without sensitivity to the microscopic details and quantify these effects using concepts from information theory.
In this way, the method facilitates the analysis of the relation between microscopic processes and a their aggregation to a macroscopic level of description and  informs about the complexity of a system introduced by heterogeneous interaction relations.
\end{abstract}

\keywords{Opinion Dynamics; Contrarian Voter Model; Markov Chains; Aggregation; Emergence; Lumpability; Agent-based models.}

\section{Introduction} 
\label{introduction}

% % REV: first paragraph rewritten to remove quotation and clarify general scope

Agent-based models (ABMs) are an attempt to understand how macroscopic regularities may emerge through processes of self-organization in systems of interacting agents.
One of the main purposes of this modeling strategy is to shed light on the fundamental principles of self-organized complexity in adaptive multi-level systems in order to gain an insight into the microscopic conditions and mechanisms responsible for the temporal and spatial patterns observed at aggregate levels. 
Therefore, ABMs are sometimes considered as a methodology to provide a >>theoretical bridge<< (\cite{Macy2002}:148) between micro and macro theories (see also \cite{Saam1999,Squazzoni2008}).

% %One of the main purposes of this modeling strategy is >>to enrich our understanding of fundamental processes<< (\cite{Axelrod1997}:25) underlying certain observed patterns, or to >>explore the simplest set of behavioral assumptions required to generate a macro pattern of explanatory interest<< (\cite{Macy2002}:146).
% %Therefore, ABMs are sometimes considered as a methodology to provide a >>theoretical bridge<< (\cite{Macy2002}:148) between micro and macro theories (see also \cite{Saam1999,Squazzoni2008}).

% % REV: This paragraph is added in order to make clear the principal motivation of the paper. Remove Banisch2012arx add Banisch2014dnc. 

While used as a tool in economics, sociology, ecology and other disciplines, ABMs are often criticized for being tractable only by simulation.
This paper addresses this issue by applying Markov chain and information-theoretic tools to a particular ABM with the specific objective to better understand the transition from the most informative agent level to the levels at which the system behavior is typically observed.
A well posed mathematical basis for linking a micro-description of a model to a macro-description may help the understanding of many of the observed properties and therefore provide information about the transition from the interaction between individual entities to the complex macroscopic behaviors observed at the global level.

% % REV: a short description of the idea of micro chains has been added because the lengthy next paragraph is skipped

For this purpose, the paper draws upon a recently introduced Markov chain framework for aggregation in agent-based and related computational models  (see \cite{Banisch2012son,Banisch2013eccs,Banisch2014dnc,Banisch2013acs} and also \cite{Banisch2014phd}).
%which makes use of lumpability theory (see \cite{Banisch2012son,Banisch2012arx,Banisch2013eccs,Banisch2013acs} and also \cite{Banisch2014phd}).
The starting point is a microscopic Markov chain description of the dynamical process in complete correspondence with the dynamical behavior of the agent model, which is obtained by considering the set of all possible agent configurations $\BSigma$ as the state space of a huge Markov chain -- an idea borrowed from \cite{Izquierdo2009}.
Namely, if we consider an ABM in which $N$ agents can be in $\delta$ different states this leads to a Markov chain with $\delta^N$ states.
Moreover, in models with sequential update by which one agent is chosen to update its state at a time, transitions are only allowed between system configurations that differ with respect to a single agent.
Such an explicit micro formulation enables the application of the theory of Markov chain aggregation -- namely, lumpability \cite{Kemeny1976} -- in order to reduce the state space of the micro chain and relate microscopic descriptions to a macroscopic formulation of interest \cite{Banisch2012son}.

% % Removed because we will derive the micro chain for the CVM later and there is probably no need to complify the introductory description with these generalizations. A short description is added above.

%Consider an ABM defined by a set $\N$ of agents, each one characterized by individual attributes that are taken from a finite list of possibilities.
%We denote the set of possible attributes by $\S$ and we call the \emph{configuration space} $\BSigma$ the set of all possible combinations of attributes of the agents, i.e. $\BSigma = \S^N$.
%Therefore, we denote an \emph{agent configuration} as $\x \in \bf{\Sigma}$ and write $\x = (x_1,\dots, x_i, \dots, x_N)$ with $x_i \in \S$.
%The updating process of the attributes of the agents at each time step typically consists of two parts. 
%First, a random choice of a subset of agents is made according to some probability distribution $\omega$. 
%Then the attributes of the agents are updated according to a rule $\u$, which depends on the subset of agents selected at this time.
%With this specification, ABMs can be represented by a so-called random map representation which may be taken as an equivalent definition of a Markov chain  \cite{Levin2009}.
%Following \cite{Banisch2012son}, we refer to the process $(\BSigma,\Phat)$ as \emph{micro chain}.

\begin{figure}[hbtp]
\centering
\includegraphics[width=0.95\linewidth]{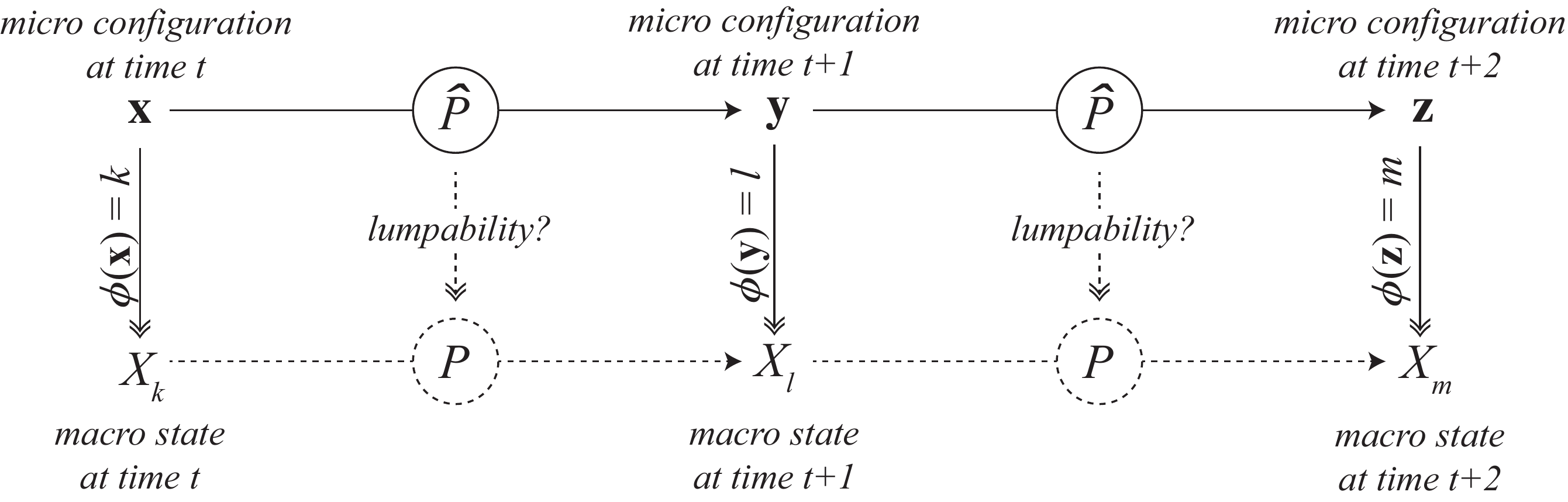}
\caption{A micro process ($\x,\y, {\bf z} \in \BSigma$) is observed ($\phi$) at a higher level and this observation defines another macro level process ($X_k,X_l,X_m \in \X$). The micro process is a Markov chain with transition matrix $\Phat$. The macro process is a Markov chain (with $P$) only in the case of lumpability.}
\label{fig:ProjectionGeneral}
\end{figure}

Namely, when performing simulations of an ABM we are actually not interested in all the dynamical details, but rather in the behavior of certain macro-level properties that inform us about the global state of the system.
In opinion dynamics, and in binary opinion models in particular, the typical level of observation is the number of agents in the different opinion states or respectively the average opinion (due to the analogy to spin systems often called >>magnetization<<).
% (such as average opinion, number of communities, etc.).
The explicit formulation of ABMs as Markov chains enables the development of a mathematical framework to link a micro chain corresponding to an ABM to such a macro-level description of interest.
More precisely, from the Markov chain perspective, the transition from the micro to the macro level is a projection of the micro chain with state space $\BSigma$ onto a new state space $\bf{X}$ by means of a (projection) map $\Pi_{\phi}$ from $\bf{\Sigma}$ to $\bf{X}$.
The meaning of the projection $\Pi_{\phi}$ is to lump sets of micro configurations in $\bf{\Sigma}$ into an aggregate set according to the macro property of interest.
Such a situation naturally arises if the ABM is observed not at the micro level of $\BSigma$, but rather in terms of a measure $\phi$ on $\BSigma$ by which all configuration in $\BSigma$ that give rise to the same measurement are mapped into the same macro state, say $X_k \in \X$.
An illustration of such a projection is provided in Fig. \ref{fig:ProjectionGeneral}.

% % REV: Extension to make clear that the Markovian and the non-Markovian case 

Two things may happen by projecting the microscopic Markov chain onto a coarser partition.
First, the macro process is still a Markov chain which is the case of lumpability \cite{Kemeny1976}.
Then Markov chain tools can be used to compute the dynamical quantities of interest and a precise understanding of the model behavior is possible.
As shown in \cite{Banisch2014dnc,Banisch2013acs} this depends essentially on the symmetries implemented in the model.
Secondly, Markovianity may be lost after the projection which means that memory effects are introduced at the macroscopic level.
Noteworthy, in ABMs as well as more generally in Markov chains, this situation is the rule rather than an exception \cite{Chazottes2003,Gurvits2005,Banisch2012son}.
While the first part of this paper derives exact Markov chain description for the complete and a perfect two--community graph, and the second part of this paper is devoted to the study of the non-Markovian case.

% % REV: This paragraph is extended in order to clarify that the explicit construction of micro chains is needed. It also contains a series of additional references to some information-theoretic approaches to quantify deviations from Markovianity

For these purposes, the explicit construction of an ABM's microscopic transition kernel is a necessary starting point because it helps, on the one hand, to establish the conditions for which the macro-level process remains Markovian (i.e., lumpability), and enables, on the other hand, the use of information-theoretic measures for the "closedness" of an aggregate description.
A series of measures -- among them conditional past-future mutual information \cite{Goernerup2008} and micro-to-macro information flow \cite{Pfante2013} -- have been developed to quantify the deviations from an idealized description which models the dynamics of the system by the state variables associated to a coarser level.
See also \cite{Shalizi2003,Jacobi2009,Goernerup2010,Pfante2014cha} for the mathematical analysis of the relation between different levels of description in complex multi-level systems.
In our context, this allows for a quantification of the memory effects that are introduced by a global aggregation over the agent population without sensitivity to micro- or mesoscopic structures which presents a first step to study how microscopic heterogeneity in ABMs may lead to macroscopic complexity when the aggregation procedure defines a non-Markovian macro process.
To my knowledge, this paper is the first to apply these concepts to an ABM.

% % Stuff
%are a promising tool to study the relationship between different levels of description in complex multilevel systems such as ABMs. 
%Global aggregation over an agent population without sensitivity to micro- or mesoscopic structures leads to memory effects at the macroscopic level.
%information-theoretic measures such as conditional past-future mutual information (intra-level deviation from Markovianity) and micro-to-macro information flow (inter-level flow) provide a reasonable framework to study these effects \cite{Banisch2014acscodym,Banisch2014eccs}.
%The method informs in this way about the complexity of a system introduced by microscopic constraints, non-trivial interaction relations and other patterns of heterogeneity at the individual level.

% % REV: added as well to better describe opinion models

%The model used to illustrate these ideas is an extension of the voter model (VM) that includes contrarian behavior.

In this paper the contrarian voter model (CVM) is used as a first simple scenario.
The CVM is a binary opinion model where $N$ agents placed on a network can adopt two different opinions: $\square$ and $\blacksquare$.
In the pure voter model \cite{Kimura1964,Clifford1973,Liggett1999} two agents are chosen at random and they align in the interaction because one of them imitates the other.
As in other binary models of opinion dynamics (see \cite{Galam2002,Sznajd-Weron2004} for two well-known variants and \cite{Castellano2009} for an overview) this mechanism of local alignment leads to a system which converges to a final profile of global conformity (consensus) in which all agent share the same opinion.
Contrarian behavior, then, relates to the presence of individuals that do not seek conformity under all circumstances or to the existence of certain situations in which agents would not desire to adopt the behavior or attitude of their interaction partner.
In the CVM studied here contrarian behavior is included by introducing a small probability $p$ with which agents to not imitate their interaction partner, but adopt precisely the opposite opinion.

There are several different ways to include nonconformity behavior into a binary opinion model and the approach adopted here is probably the most simple one.
In our choice, we basically follow Ref. \cite{Galam2004} which, based on the concept of contrarian investment strategies in finance \cite{Dreman1980,Corcos2002}, is the first study to introduce contrarian behavior into a model of opinion dynamics (namely, into Galam's majority model).
While the majority model without contrarians is characterized by a relatively fast convergence to complete consensus, the introduction of only a small rate of contrarian choices leads to the coexistence of the two opinions with a clear majority-minority splitting. 
Noteworthy, as the contrarian rate increases further, the model exhibits a phase transition to a disordered phase in which no opinion dominates in the population.
Similar observations have been made for the Sznajd model \cite{Lama2005,Sznajd-Weron2011,Nyczka2012}.

More recently the literature often distinguishes between two types of nonconformity: (i.) anti-conformity or contrarian behavior and (ii.) independent or inflexible agents \cite{Sznajd-Weron2011,Nyczka2012,Crokidakis2014}.
See \cite{Mobilia2003,Galam2007,Sznajd-Weron2011,Nyczka2012,Crokidakis2014,Maity2014} for opinion models that include independent or inflexible agents.
The importance of a distinction between individuals that generally oppose the group norm or act independently of it is, from the socio-psychological perspective, relatively obvious.
The fact that these two behaviors may also give rise to qualitatively different dynamical properties, however, has been established only recently \cite{Nyczka2012}.
Here we stick to contrarians.

The voter model with contrarians presented in \cite{Masuda2013} is probably the one that relates most to the model used here.
The main difference is that a fixed number of agents always acts in a contrarian way whereas in the present model all agents take contrarian choices with a small probability $p$.
In that setting, \cite{Masuda2013} could not observe the phase transition from majority-minority splitting to disorder, but rather a change from a uniform to a Gaussian equilibrium distribution.
While this difference in comparison with the Sznajd and Galam models has been attributed to the linearity of the CVM in \cite{Masuda2013}, this study shows that there is in fact an order-disorder phase transition in the CVM as well.
However, the ordered phase can be observed only below a very small contrarian rate of $p^* = 1/(N+1)$ at which the equilibrium distribution is uniform (see Sec. \ref{cha:5.HMStatDyn}), in accordance with \cite{Masuda2013}.
In the setting of \cite{Masuda2013} with a fixed number of contrarian agents, however, this value is already reached, on average, with only a single contrarian, independent of the population size.

Notice, finally, that the complete graph plays an exceptional role for the analytical treatment of nonconformity models commended on above.
From the Markovian point of view, this is due to the fact that for the complete graph binary opinion models are lumpable -- that is reducible without loss of information -- to a macroscopic description in terms of the average opinion or >>magnetization<< \cite{Banisch2012son}.
This paper analyses the complete graph as well, but it goes beyond it by studying the CVM on a perfect two--community graph.
In that case, a loss-less macro description is obtained by taking into account separately the average opinion in the two sub-graphs, that is, by a refinement of the level of observation.
Alongside with the analysis of the respective model dynamics, these two cases allow to address some very interesting questions concerning the relation between the two coarse-grainings.
For instance, it is possible to illustrate why lumpability (in its strong as well as in its weak form) fails for the two--community graph.
Moreover, looking at the two-community scenario from the global perspective of total >>magnetization<< allows for the exact computation of memory effects that emerge at the macroscopic level.

% % Conclusion

%In comparison with the Sznajd and Galam models, the CVM does not lead to stable co-existence patterns in which the minority persists, but rather in a 
%The ultimate and broader goal of this enterprise is to better understand the microscopic conditions for long range correlations .

%The objective of this paper is two-fold.
%On the one hand, it presents an application of these theoretical ideas to a simple opinion model with contrarian behavior
%-- the contrarian voter model (henceforth CVM) -- including a Markov chain solution for homogeneous mixing and the two-community model.
%Contrarian behavior relates to the presence of individuals that do not seek conformity under all circumstances or to the existence of certain situations in which agents would not desire to adopt the behavior or attitude of their interaction partner.
%Contrarian behavior has been addressed previously, see, for instance, \cite{Galam2004,Li2012}.
%On the other hand, the paper aims to go beyond Markovian aggregation by considering the case that an observation or aggregation fails to define a Markovian macro process.
%Noteworthy, in ABMs as well as more generally in Markov chains, this situation is the rule rather than an exception \cite{Chazottes2003,Gurvits2005,Banisch2012son}.
%Using the CVM as an example, the paper presents a first step to study how microscopic heterogeneity in ABMs may lead to macroscopic complexity when the aggregation procedure defines a non-Markovian macro process.

% % % % REV01

The sequel of the paper is organized as follows.
Sec. \ref{cha:5.CVM} introduces the CVM and derives the corresponding microscopic Markov chain. 
Sec. \ref{cha:5.HMand2Com} deals with the model on the complete and the two-community graph with a particular focus on the stationary dynamics of the model. 
%Homogeneous mixing leads to a random walk on the line with $N+1 = O(N)$ states whereas the two-community model to a random walk on a 2D lattice with $O(N^2)$ states.
The model dynamics are studies in terms of the contrarian rate $p$ and the coupling $r$ between the two-communities.
After the discussion of the two stylized topologies, Sec. \ref{cha:5.NetworkDynamics} shows the effect of various paradigmatic networks on the macroscopic stationary behavior.
In Sec. \ref{cha:5.MicroMesoMacro} we return to the two-community CVM and find with it an analytical scenario to study the discrepancy between a mean-field model (homogeneous mixing) and the model on a more complex (though still very simple) topology.
It shows that memory effects are introduced at the macro level when we aggregate over agent attributes without sensitivity to the microscopic details.

%\section{Markov Chain Aggregation}

% micro and macro perspective
% definitions, configurations,
% single-step dynamics

\section{The Contrarian Voter Model}
\label{cha:5.CVM}

%In order to address questions related to the macro effects of heterogeneous interaction probabilities we will concentrate on an extension of the VM called contrarian voter model (CVM) or sometimes anti-voter model.
%Contrarian behavior relates to the presence of individuals that do not seek conformity under all circumstances or to the existence of certain situations in which agents would not desire to adopt the behavior or attitude of their interaction partner.
%Contrarian behavior has been addressed previously, see, for instance, \cite{Galam2004,Li2012}.

\subsection{Model}

The CVM is a binary opinion model where $N$ agents can adopt two different opinions: $\square$ and $\blacksquare$.
The model is an extension of the voter model (VM) in order to include a form of contrarian behavior. %, that is, 
At each step, an agent ($i$) is chosen at random along with one of its neighbors ($j$).
Usually (with probability $1-p$), $i$ imitates $j$ (VM rule), but there is also a small probability $p$ that agent $i$ will do the opposite (contrarian extension).
More specifically, if $i$ holds opinion $\blacksquare$ and meets an agent $j$ in $\square$, $i$ will change to $\square$ with probability $1-p$, and will maintain its current state $\blacksquare$ with probability $p$.
Likewise, if $i$ and $j$ are in the same state, $i$ will flip to the opposite state with probability $p$.

% % REV: removed paragraph and table

\rem{
While the VM rule may be interpreted as a kind of ferromagnetic coupling by which neighboring spins (agents) align, the contrarian rule can be interpreted as anti-ferromagnetic coupling by which neighbors are of opposed sign after the interaction.
Table \ref{tab:5.1} illustrates the CVM update rules where $x_i,x_j$ denotes the current states of agents $i$ and a neighbor $j$ and $y_i$ the updated state at time $t+1$.

\begin{table}[h]
\centering
\begin{tabular}{| c | c c |}
\hline
Prob.			 	& $x_j$ & $x_j$ \\\hline
$x_i$			& $y_i$ & $y_i$\\
$x_i$			& $y_i$ & $y_i$\\\hline
\end{tabular}
\begin{tabular}{| c | c c |}
\hline
$(1-p)$			 			& $\blacksquare$ & $\square$ \\\hline
$\blacksquare$ 	& $\blacksquare$ & $\square$\\
$\square$ 			& $\blacksquare$ & $\square$\\\hline
\end{tabular}
\begin{tabular}{| c | c c |} 
\hline
$p$		 			& $\blacksquare$ & $\square$ \\\hline
$\blacksquare$ 	& $\square$ & $\blacksquare$\\
$\square$ 			& $\square$ & $\blacksquare$\\\hline
\end{tabular}
\caption{Update rules $y_i = \u(x_i,x_j)$ for the CVM. The VM rule (ferromagnetic coupling) is applied with probability $(1-p)$, the contrarian rule (anti-ferromagnetic coupling) with probability $p$. }
\label{tab:5.1}
\end{table}
}

\subsection{Markovian Micro Dynamics}

From the micro-level perspective, the CVM implements an update function of the form $\u : \S \times \S \times \Lambda \rightarrow \S$.
For further convenience, we denote the current state of an agent $i$ by $x_i$ and the updated state at time $t+1$ as $y_i$.
The map $\u$ can then be written as
%That is, the new state of a randomly chosen agent $i$, denoted as $y_i$, is given by
\begin{eqnarray}
y_i = \u(x_i,x_j,\lambda) =
 \left\{
 \begin{array}{c l}
  x_j & : \lambda = \lambda_V\\
 \bar{x}_j & : \lambda = \lambda_C
 \end{array}\right\},
 \label{eq:VMContrarian}
\end{eqnarray}
where $\bar{x}_j$ denotes the opposite attribute of $x_j$.
In each iteration, two agents $i,j$ are chosen along with a random variable $\lambda \in \Lambda = \{\lambda_V,\lambda_C\}$ that decides whether the voter ($\lambda_V$) or the contrarian rule ($\lambda_C$) is performed.
The probability for that is $\omega(i,j,\lambda)$.
Notice that the update rule is equal for all agents and independent from the agent choice. 
Therefore the probability that an agent pair $(i,j)$ is chosen to perform the contrarian rule can be written as $\omega(i,j) Pr(\lambda = \lambda_C) = p \omega(i,j)$.
Respectively, we have $(1-p) \omega(i,j)$ for the VM rule.
Moreover, the respective probabilities $\omega$ are equal for each step of the iteration process.
Consequently, the CVM iteration may be seen as a (time-homogeneous) random choice among deterministic options, and is therefore a Markov chain (see \cite{Levin2009,Banisch2012son} and \cite{Banisch2014phd} for all details).

Due to the sequential update scheme only one agent (namely, agent $i$) may change its state at each time step.
This means that a non-zero transition probabilities exist only between agent configurations $\x, \y$ that differ in at most one element (agent). 
We call those configuration adjacent and write $\x \stackrel{i}{\sim} \y$ if $x_i \neq y_i$ and $x_k = y_k \forall k \neq i$.
Considering that $\lambda_C$ (for contrarian rule) is chosen with probability $p$ and $\lambda_V$ (VM rule) with $(1-p)$, and that this choice is independent of the agent choice, the micro-level transition probability $\Phat(\x,\y)$ between two adjacent configurations $\x \stackrel{i}{\sim}  \y$ is given by
\begin{equation} 
\Phat(\x,\y) = (1-p) \sum\limits_{j: \left(\substack{y_i = x_j} \right) }^{} \omega(i,j) + p \sum\limits_{j: \left(\substack{y_i = \bar{x}_j} \right) }^{} \omega(i,j).
\label{eq:PhatVMContrarian}
\end{equation}
Notice that the configuration space of the CVM ($\BSigma$) is the set of all bit-strings of length $N$, and that therefore, the micro-level process for the CVM corresponds to a random walk on the hypercube (as for the VM, see \cite{Banisch2012son,Banisch2014dnc}).
Notice also that the CVM leads to a \emph{regular} chain (as opposed to an absorbing random walk for the original VM) because whenever $p > 0$, there is a non-zero probability that the process leaves the consensus states $(\square\square\ldots\square)$ and $(\blacksquare\blacksquare\ldots\blacksquare)$.
Eq. (\ref{eq:PhatVMContrarian}) tells us that this probability is precisely $p$.
Therefore, the system does not converge to a fixed configuration and the long-term behavior of the model can be characterized by its stationary distribution.

\subsection{Full Aggregation}
\label{sec:2.FullAgg}

% % REV: condense

The most natural level of observation in binary state dynamics is to consider the temporal evolution of the attribute densities, or respectively, the number of agents in the two different states.
While a mean-field description would typically formulate the macro dynamics as a differential equation describing the evolution of attribute densities, the Markov chain approach operates with a discrete description (in time as well as in space) in which all possible levels of absolute attribute frequencies and transitions between them are taken into account.

Let us denote as $k$ the number of $\square$-agents in the population ($k = N_{\square}$) and refer to this level of observation as global or \emph{full aggregation}.
In terms of Markov chain aggregation \cite{Banisch2012son}, a macro description in terms of $k$ is achieved by a projection of the micro-level process on the hypercube onto a new process with the state space defined by the partition $\X = \{X_0,\ldots,X_k,\ldots,X_N\}$, where $0 \leq k \leq N$.
Notice that in hypercube terminology that level of observation corresponds to the Hamming weight of a configuration $h(\x) = k$ and each $X_k$ collects all micro configurations with the same Hamming weight.

%Regardless of the microscopic details such as more complex interaction networks or rules, a macro level description of that kind, which is always a tremendous reduction of original system, is desirable in order to obtain a better understanding of the model behavior.
%As a matter of fact, it is desirable for both numerical as well as analytical arguments.

%One of the main contributions of the framework proposed in \cite{Banisch2012son} is that the link between the microscopic system and a certain macro level description is made explicit.
%Namely, a system property induces a partition on the space of all possible micro configurations and the macro-level process corresponds to the micro process projected onto that partition.
%For the CVM and many other binary models the most relevant macro formulation corresponds to a projection of the micro-level process on the hypercube onto a macro-level process with the state space defined by the partition $\X = \{X_0,X_1,\ldots,X_k,\ldots,X_N\}$, where $0 \leq k \leq N$ corresponds to the number of $\square$-agents in the population.
%Notice that in hypercube terminology that level of observation corresponds to the Hamming weight of a configuration $h(\x) = k$ and each $X_k$ collects all micro configurations with the same Hamming weight.

\begin{figure}[hbt]
\centering
\includegraphics[width=1.0\linewidth]{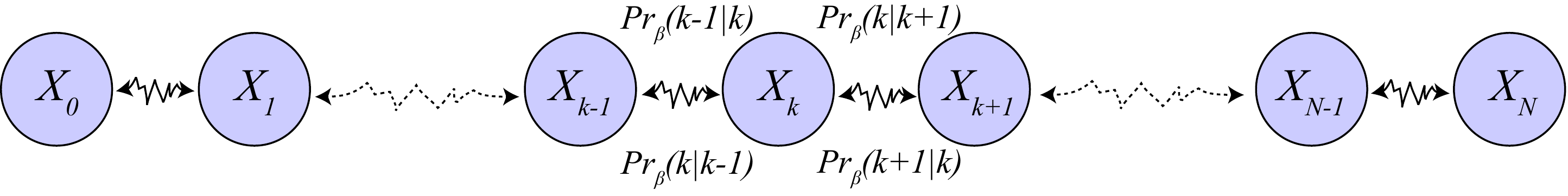}
\caption{Full aggregation is obtained by the agglomeration of states with the same Hamming weight $h(\x) = k$. The resulting macro process is, in general, a non-Markovian process on the line.}
\label{fig:FullAggregation}
\end{figure}

%In this regard, o
One important observation in \cite{Banisch2012son,Banisch2013acs} has been that homogeneous mixing is a prerequisite for lumpability with respect to $\X$, and that microscopic heterogeneities (be it in the agents or in their connections) translate into dynamical irregularities that prevent lumpability.
This means that full aggregation over the agent attributes ($k = h(\x)$) leads in general to a non-Markovian macro process.
We illustrate this process in Fig. \ref{fig:FullAggregation}.
Still, the process obtained by the projection from micro to macro is characterized by the fact that from an atom $X_k$ the only possible transitions are the loop to $X_k$, or a transition to neighboring atoms $X_{k-1}$ and $X_{k+1}$ because 
%This is of course due to the fact that the CVM implements single-step dynamics in which 
only one agent changes at a time.
However, the micro level transition rates (\ref{eq:PhatVMContrarian}) depend essentially on the connectivity structure between the agents, and therefore, the transition probabilities at the macro level (denoted as $Pr_{\beta}(l|k)$ in Fig. \ref{fig:FullAggregation}) are not uniquely defined (except for the case of homogeneous mixing).
That is, for two configurations in the same macro state $\x,\x' \in X_k$ the probability to go to another macro state (e.g., $X_{k+1}$) may be very different which violates the lumpability conditions of Thm. 6.3.2 in \cite{Kemeny1976}.
Before we use the CVM to address questions related to non-lumpable projections, we discuss the model behavior using Markov chain tools for two stylized situations.

%Later in this paper, we use the CVM as a very simple model in which questions related to non-lumpable projections can be addressed.

%For instance, why and in what sense does the behavior of the macro process deviate from Markovianity? 
%Do we introduce memory or long-range correlations at the macro level by the very way we observe the process? 
%The scope is rather theoretical in the sense that we are interested in the kind of macro effects to which a non-trivial interaction structure at the micro level may lead.
%The questions we aim to address are of the following type:
%Is the emergence of these effects just due to an aggregation which is insensitive to microscopic heterogeneities?

\section{Homogeneous Mixing and the Two-Community Graph}
\label{cha:5.HMand2Com}

This section analyses the behavior of the CVM for homogeneous mixing and the two-community graph.
As shown in \cite{Banisch2013acs} at the example of the VM, the symmetries in the interaction network can be used to define a Markovian coarse-graining.
As both interaction topologies are characterized by a large symmetry group, Markov chain aggregation considerably reduces the size of the micro chains such that the important entities of interest (e.g., stationary distribution) can be computed on the basis of the respective macroscopic transition matrices.

\subsection{Homogeneous Mixing}

% % REV: strong reduction in order to simplify the description. Mainly, some details of the formalism are skipped as the way in which the resulting macro description is obtained is visible without these details.

The case of homogeneous mixing is particularly simple.
Let us consider that the model is implemented on the complete graph without loops where the probability to choose a pair $(i,j)$ of agents becomes $\omega(i,j) = 1/N(N-1)$ whenever $i \neq j$ and $\omega(i,i)=0, \forall i$.
%This means that the interaction structure is invariant with respect to all agent permutations (that is, $\omega(i,j) = \omega(\sigma i, \sigma j),\forall \sigma \in \SG_N$ and all pairs $(i,j)$) and \cite{Banisch2013acs}, Prop. 1, tells us that all agent configurations with the same number $k$ of agents in $\square$ (and therefore $N-k$ in $\blacksquare$) belong to a class of macroscopic equivalence and can be mapped into the same macro atom ($X_k$).
%In other words, for homogeneous mixing \emph{full aggregation} over all agents does not destroy Markovianity, which is in complete analogy to the pure VM.
Consequently, since all agents interact with all the others with equal probability, the respective transition rates depend only on the numbers $k$ and $N-k$ of agents in the two states: 
%For instance, $P(X_k,X_{k+1})$ can be obtained by evaluating the transition probability (\ref{eq:PhatVMContrarian}) from some $\x \in X_k$ to the set of all $\y \in X_{k+1}$.
%Then we obtain 
\begin{eqnarray} 
 \begin{array}{l l}
P(X_k,X_{k+1}) %&= \sum\limits_{\y \sim \x} \Phat(\x,\y) \\
 & = \sum\limits_{x_i = \blacksquare} \left[ (1-p) \sum\limits_{j: \left(\substack{x_j = \square} \right) }^{} \omega(i,j) + p \sum\limits_{j: \left(\substack{x_j = \blacksquare} \right) }^{} \omega(i,j) \right] \vspace{6pt}\\
 & = (N-k) \left[ (1-p) k \omega + p (N-k)\omega  \right]\vspace{6pt}\\
 &= (1-p) \frac{(N-k) k}{N(N-1)} + p \frac{(N-k)(N-k-1)}{N(N-1)}.
 \end{array}
\label{eq:PMacroVMContrarian01}
\end{eqnarray}
Similarly, we obtain for $P(X_k,X_{k-1})$
\begin{eqnarray} 
\begin{array}{l l}
P(X_k,X_{k-1}) = (1-p) \frac{(N-k) k}{N(N-1)} + p \frac{k(k-1)}{N(N-1)},
\end{array}
\label{eq:PMacroVMContrarian02}
\end{eqnarray}
and finally,
\begin{eqnarray} 
\begin{array}{l l}
P(X_k,X_{k}) = %\frac{ 2k^2 + N^2 - 2 k N -N - p (N-2k)^2 }{N(N-1)}.
\frac{k^2 (2-4p)+2kN (2p - 1)+ N(N-Np+p-1)}{N (N-1)}.
\end{array}
\label{eq:PMacroVMContrarian03}
\end{eqnarray}

\begin{figure}[htbp]
	\centering
	\begin{tabular}{c c}
\includegraphics[width=0.48\textwidth]{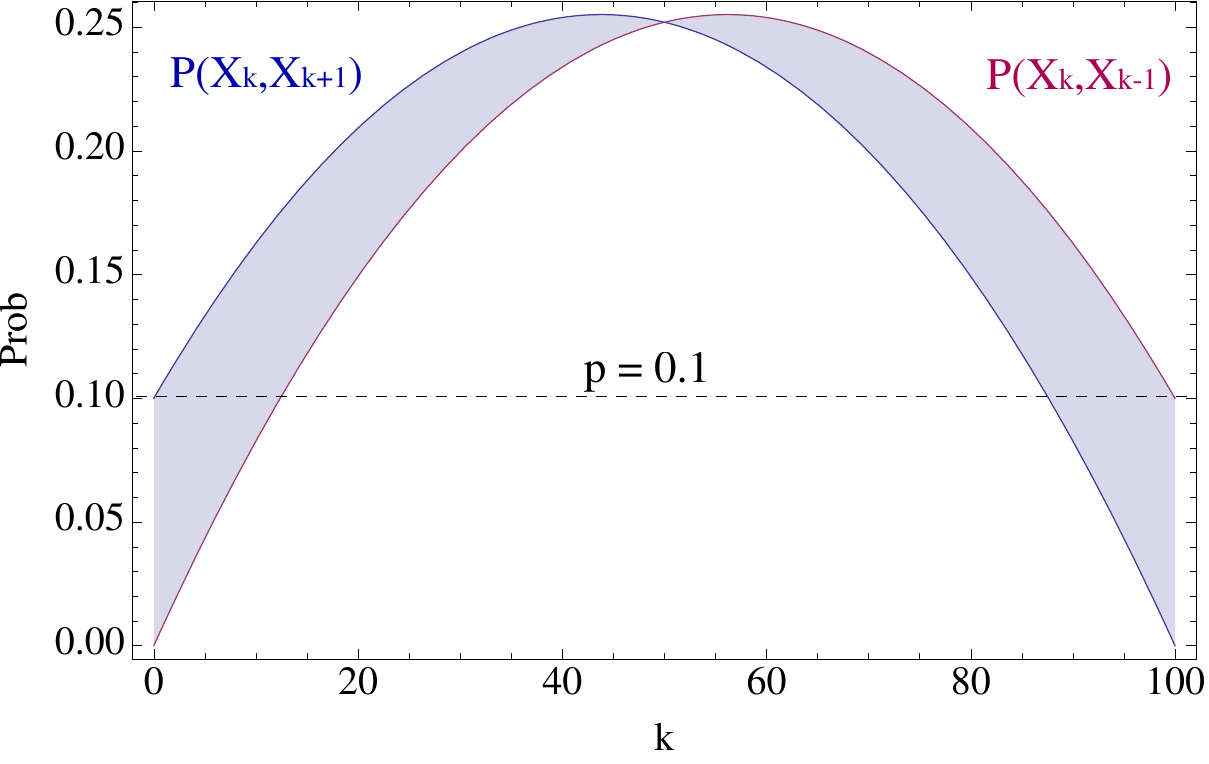}&\includegraphics[width=0.48\textwidth]{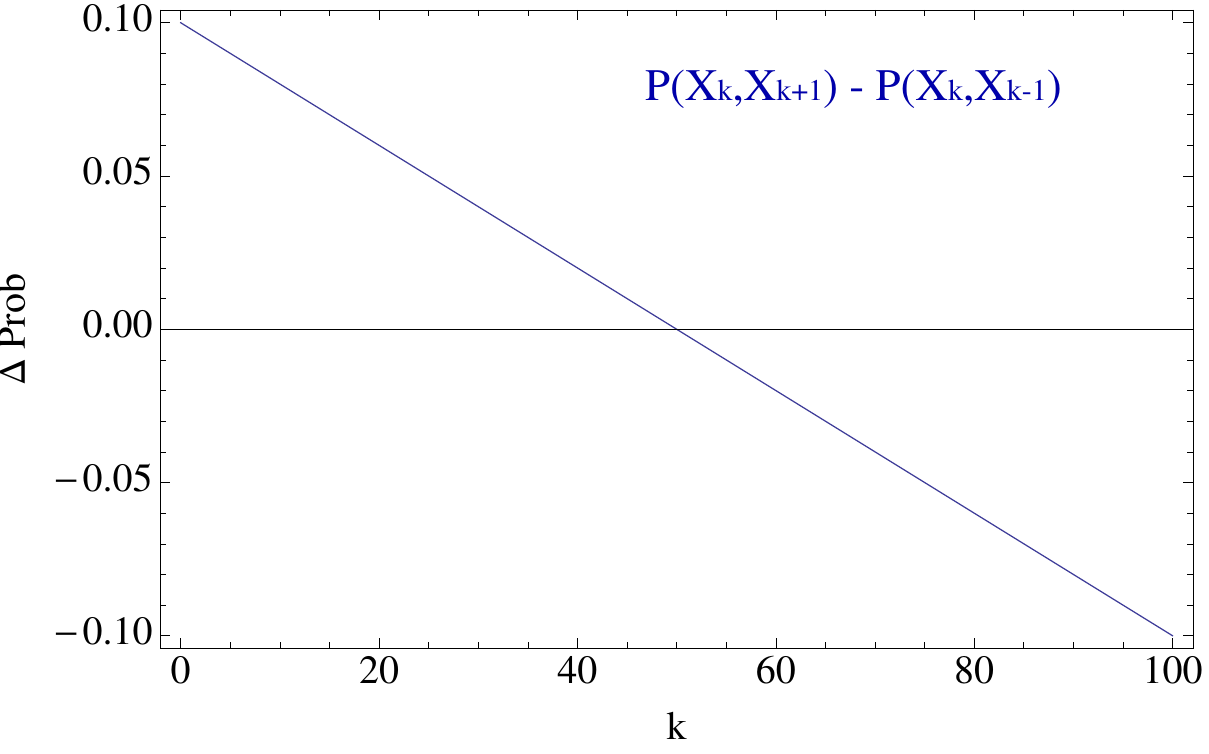}
	\end{tabular}
	\caption{Transition probabilities and difference in transition probabilities as function of $k$ ($N=100$).}
	\label{fig:CVM.Probs}
\end{figure}

Fig. \ref{fig:CVM.Probs} aims at giving an intuition about the dynamical structure of the process by considering the relation between the probability for a transition one step to the right, $P(X_k,X_{k+1})$, and a transition to the left, $P(X_k,X_{k-1})$, as a function of $k$.
This informs us about the more probable tendency for future evolution for every atom in the macro chain.
Fig. \ref{fig:CVM.Probs} shows that $P(X_k,X_{k+1}) > P(X_k,X_{k-1})$ for $k < N/2$ and respectively $P(X_k,X_{k+1}) < P(X_k,X_{k-1})$ for $k > N/2$ telling us that the contrarian rule (performed with probability $p$) introduces in every atom $X_k$ a small bias that drives the system towards the fifty-fifty configurations.
This bias is given by 
\begin{equation}
%P(X_{k+1}|X_k) - P(X_{k-1}|X_k) =
P(X_k,X_{k+1}) - P(X_k,X_{k-1}) =  p -\frac{2 k p}{N}.
\end{equation}

\subsection{Stationary Dynamics for Homogeneous Mixing}
\label{cha:5.HMStatDyn}

For homogeneous mixing, the CVM leads to a regular Markov chain on the line $\X = (X_0,\ldots,X_k,\ldots, X_N)$. 
%As already mentioned, the CVM does no longer lead to an absorbing Markov chain, but results in a regular chain.
In the case that the population reaches consensus ($k = 0$ or $k=N$) there is still a small probability, (namely $P(X_0,X_1) = P(X_{N},X_{N-1}) = p$) with which the consensus configuration is left (see Fig. \ref{fig:CVM.Probs}).
A statistical understanding of the model behavior is provided by the stationary distribution it converges to.
That is, by the distribution $\pi$ that remains unchanged under further application of $P$: $\pi P = \pi$.
%\begin{equation}
%\pi P = \pi.
%\label{eq:StatDist}
%\end{equation}
That is, the stationary distribution $\pi$ of a Markov chain $(\X,P)$ is proportional to the left eigenvector of $P$ associated to the maximal eigenvalue $\lambda_{max} = 1$ and it is well-known (e.g., \cite{Kemeny1976}:69ff) that regular chains have a unique limiting vector to which the process converges for any initial distribution.
Notice that the rate of convergence is usually related to the second largest eigenvalue of $P$ ($\lambda_2 < 1$) in the sense that the order of convergence is proportional to $\lambda_2^t$ (\cite{Kemeny1976,Behrends2000} among others).

\begin{figure}[htbp]
	\centering
		\includegraphics[width=0.49\linewidth]{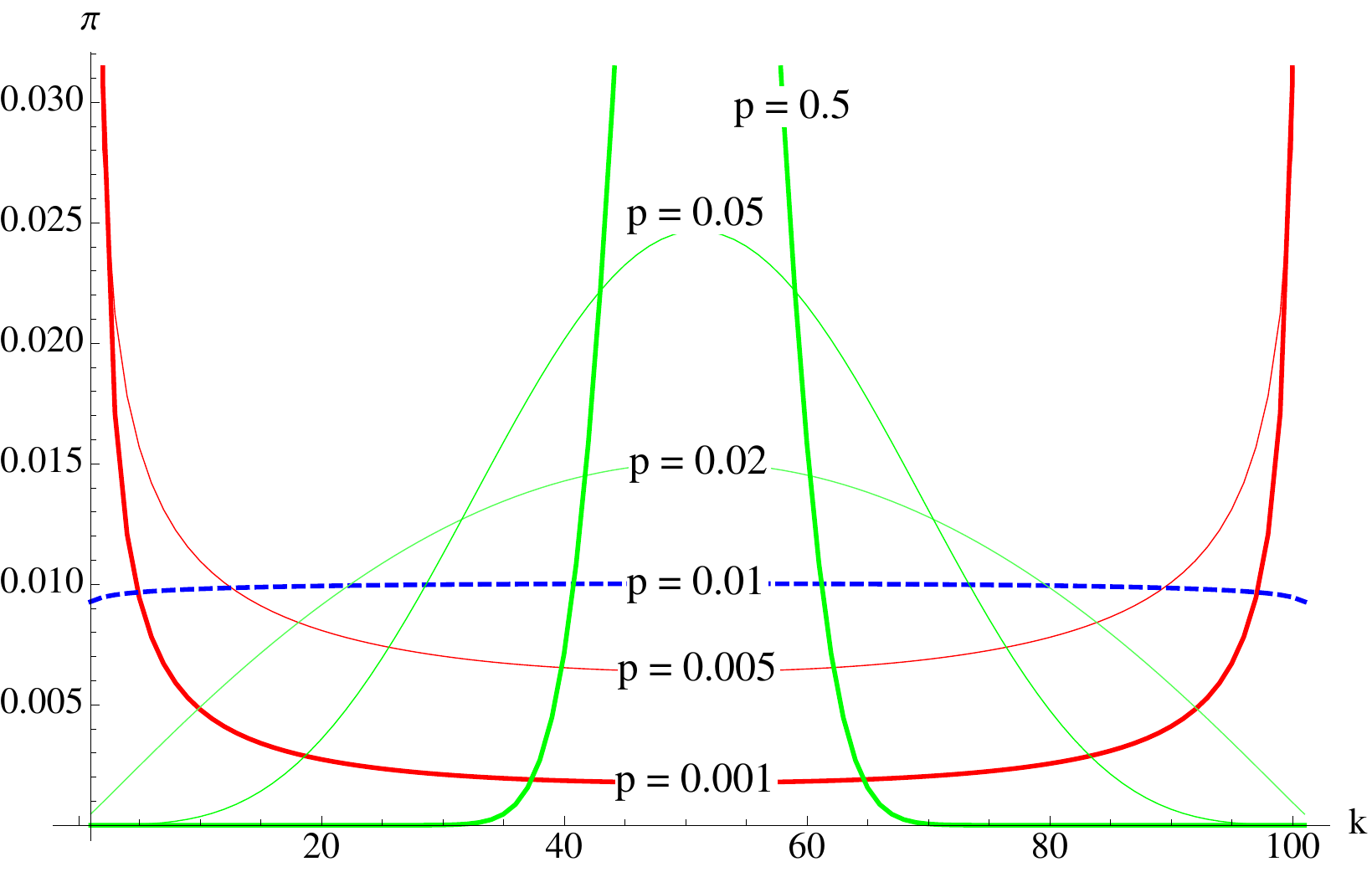}
		\includegraphics[width=0.49\linewidth]{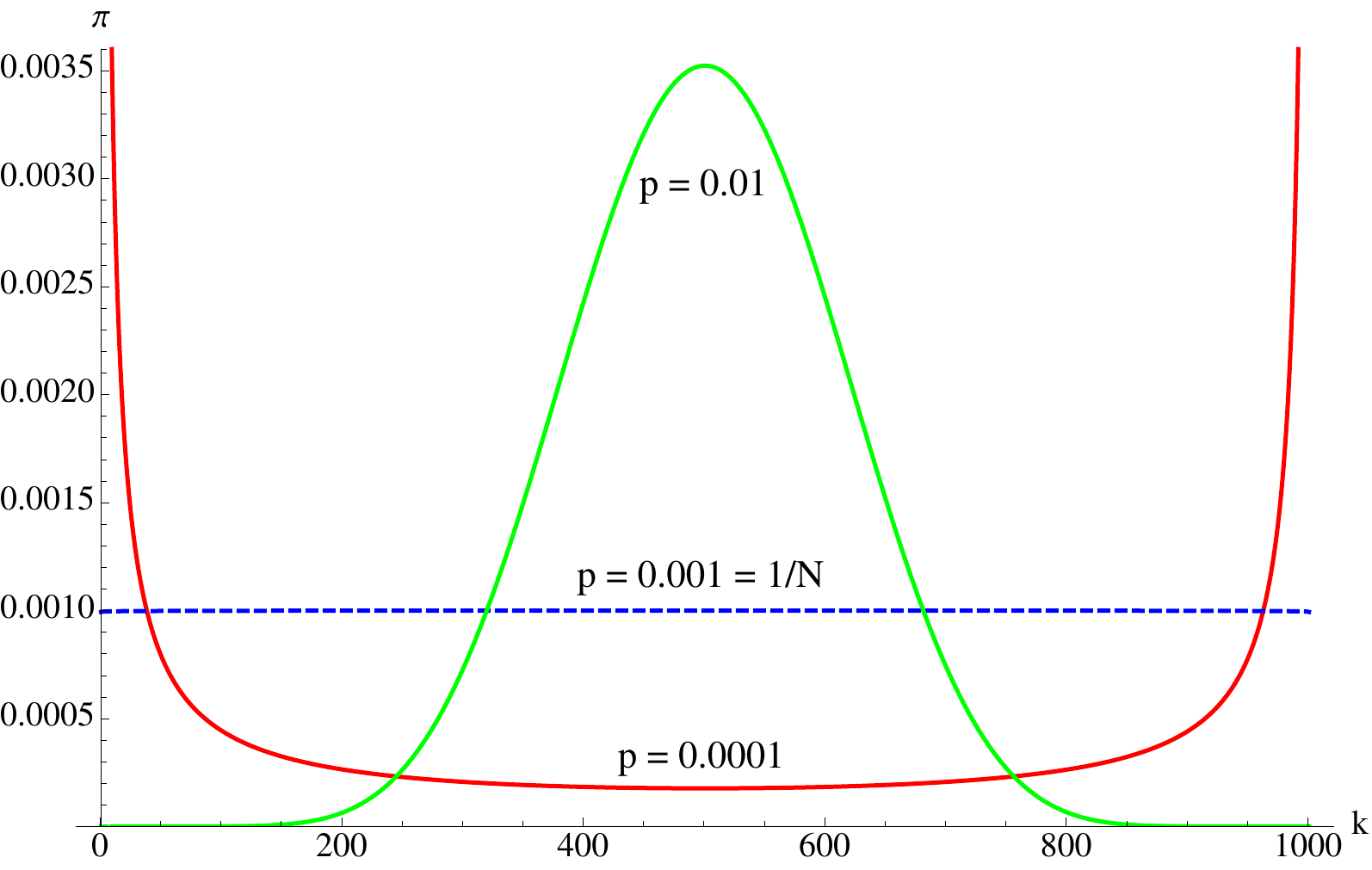}\\
		\includegraphics[width=0.80\linewidth]{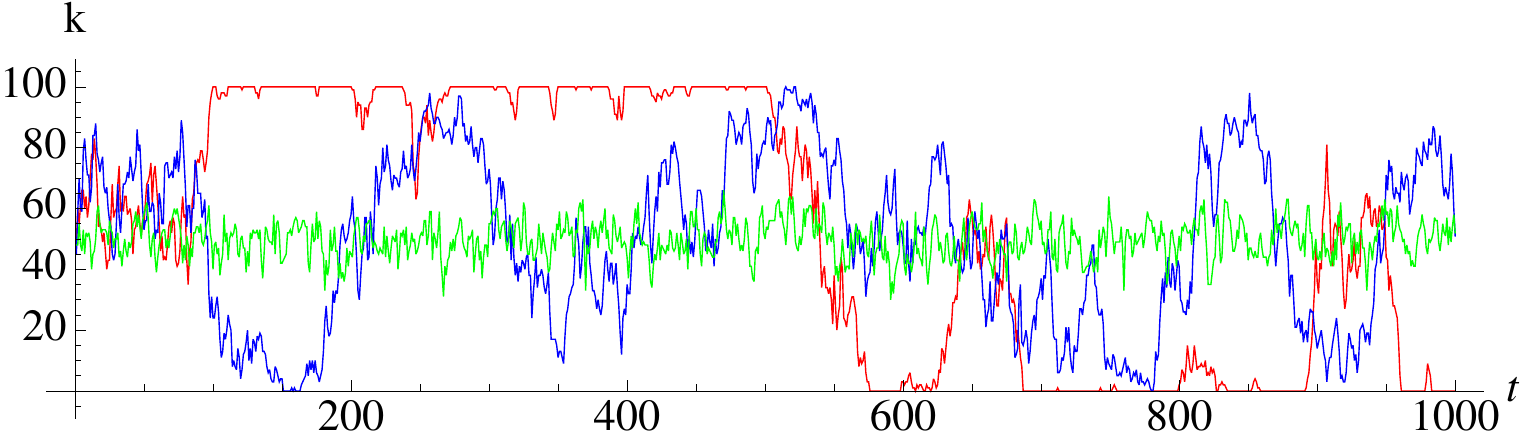}
	\caption{Stationary vector of the CVM with $N=100$ (l.h.s.) and $N = 1000$ (r.h.s.) and homogeneous mixing for various $p$. There is a transition from the absorbing VM to the random fluctuations around the mean. On the bottom, the respective example time series ($N = 100$) are shown.}
	\label{fig:CVM.StationarDist}
\end{figure}

For a system of 100 agents (respectively 1000 agents in the upper right plot) the stationary vector $\pi$ is shown in Fig. \ref{fig:CVM.StationarDist} for various contrarian rates $p$.
The horizontal axis represents the macro states $X_k$ for $k = 0,1,\ldots,N$ and the $\pi_k$ correspond to the probability with which the process is in atom $X_k$ provided it is run long enough and has reached stationarity.
Notice that, for a large number of steps, the $\pi_k$ also represent the expected value for the fraction of time the process is in $X_k$ (see Sec.4.2 in \cite{Kemeny1976}).
On the bottom of Fig. \ref{fig:CVM.StationarDist}, three characteristic time series (three single simulation runs) are shown, one for large, one for intermediate and one for low $p$ values.
This provides a better understanding of the meaning of the stationary vector in relation to the time evolution of the respective processes.

%\begin{figure}[htbp]
%	\centering
%		\includegraphics[width=0.70\linewidth]{CVM.pi.N100.pdf}\\
%		\includegraphics[width=0.70\linewidth]{CVM.timeEvolution.pdf}
%	\caption{Stationary vector of the CVM with $N=100$ and homogeneous mixing for various $p$. There is a transition from the absorbing VM to the random fluctuations around the mean. On the bottom, the respective example time series are shown.}
%	\label{fig:CVM.StationarDist}
%\end{figure}

Two different regimes can be observed in Fig. \ref{fig:CVM.StationarDist} characterized by the green and the red curves respectively.
A large contrarian rate $p$ (green curves) leads to a process which fluctuates around the states with approximately the same number of black and white agents -- the fifty-fifty situation ($k = N/2$) being the most probable observation.
The larger $p$, the lower the probability to deviate strongly from the fifty-fifty configurations.
In fact, the process resembles a random process in which agent states are flipped at random.

A different behavior is observed if $p$ is small.
This is represented by the red curves.
For a small contrarian rate, the population is almost uniform (consensus) for long periods of time, but due to the random shocks introduced by the contrarian rule there are rare transitions between the two extremes.
This is very similar to the VM at low (but non-zero) temperature, where random state switches or excitations take the role of mutations and prevent the system from complete freezing to the zero-temperature ground state.
In between these two regimes, there is a $p \approx 0.01$ for which the process wanders through the entire state space, in such a way that the stationary distribution is almost uniform.
Notice that in the case of $N = 1000$ (upper right plot), an almost uniform stationary distribution is observed for $p \approx 0.001 = 1/N$.

To be more precise, it is, in fact, not difficult to show that for any system size $N$ the stationary distribution is uniform with $\pi_k = 1/(N+1), \forall k$ exactly for $p^* = 1/(N+1)$.
All that is necessary in order to verify this is to show that $\pi P = \pi$ in this case.
Hence, we have to show that
\begin{equation}
\frac{1}{N+1} \left( P(X_{k-1},X_k) + P(X_{k},X_k) + P(X_{k+1},X_k)\right) = \frac{1}{N+1}
\label{eq:UniStatProof01}
\end{equation}
which is satisfied whenever
\begin{equation}
P(X_{k-1},X_k) + P(X_{k},X_k) + P(X_{k+1},X_k) = 1.
\label{eq:UniStatProof02}
\end{equation}
Notice that  Eq. (\ref{eq:UniStatProof02}) is equivalent to requiring that $P$ is a doubly stochastic matrix, and it is well-known that any doubly stochastic matrix has a uniform stationary vector.
It is easy to show that for the CVM, Eqs. (\ref{eq:UniStatProof01}) and (\ref{eq:UniStatProof02}) are satisfied precisely for $p^* = 1/(N+1)$, but not for other contrarian rates.

When the contrarian rate $p$ crosses the critical value $p^* = 1/(N+1)$, the system undergoes a continuous phase transition from majority-minority switching (ordered phase) to a balanced fifty-fifty situation in which no stable majorities form (disordered phase).
The fact that $p^* = 1/(N+1)$ leads to $\pi_k = 1/(N+1), \forall k$ shows the existence of large fluctuations at the critical contrarian rate, because the only way to have a stationary uniform distribution is to have very large fluctuations at any value of the state space.
For large $p$, the system behaves around the mean value (here 50 and respectively 500) with only small deviations. 
For small $p$ closed to 0, the system is rarely far from the two states of complete order (the consensus states) and in the limit of $p = 0$ has no asymptotic fluctuations at all.

% % REV: relation to previous studies

The emergence of a phase transition in the presence of contrarians has been reported in several previous studies \cite{Galam2004,Lama2005,Sznajd-Weron2011,Nyczka2012}.
However, in the Galam as well as the Sznajd model the ordered phase is characterized by the co-existence of majority and minority opinions whereas in our case the system permanently switches between the two consensus profiles.
In relation to Ref. \cite{Masuda2013}, where a fixed number of agents that always act as contrarians is used, the result derived here explains why such a phase transition cannot be observed in this setting.
The reason is that the transition value $p^* = 1/(N+1)$ scales inversely with the population size in such a way that, independent of the system size, the effective contrarian rate is already larger than $p^*$ even if there is only a single contrarian agent (in that case $p = 1/N$).
On the other hand, the observation from \cite{Masuda2013} that a single contrarian leads to a uniform stationary distribution is confirmed.

\subsection{Rate of Majority-Minority Switching}
\label{cha:5.MajMinSwitching}

\begin{figure}[ht]
	\centering
		\includegraphics[width=0.49\linewidth]{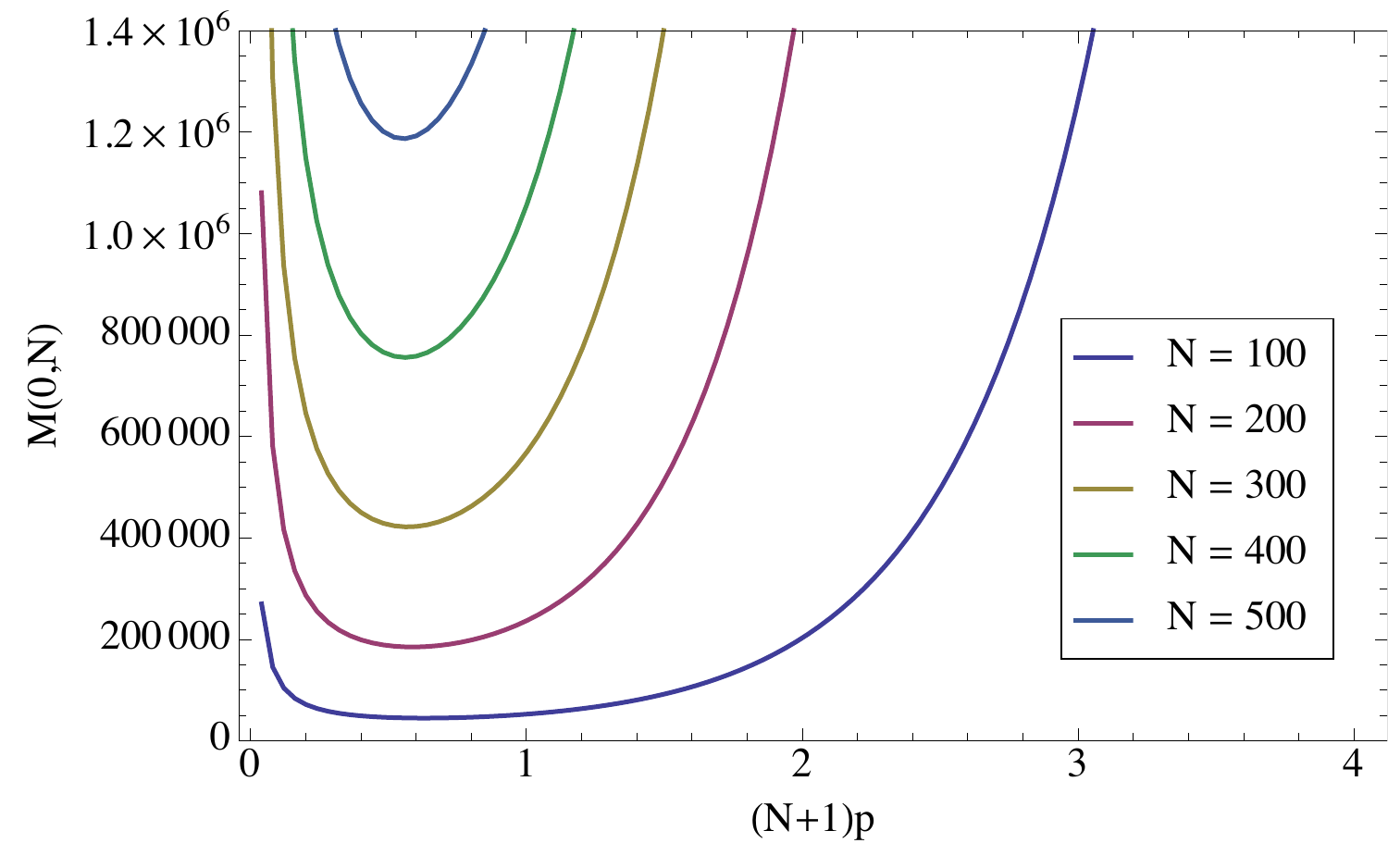}
		\includegraphics[width=0.49\linewidth]{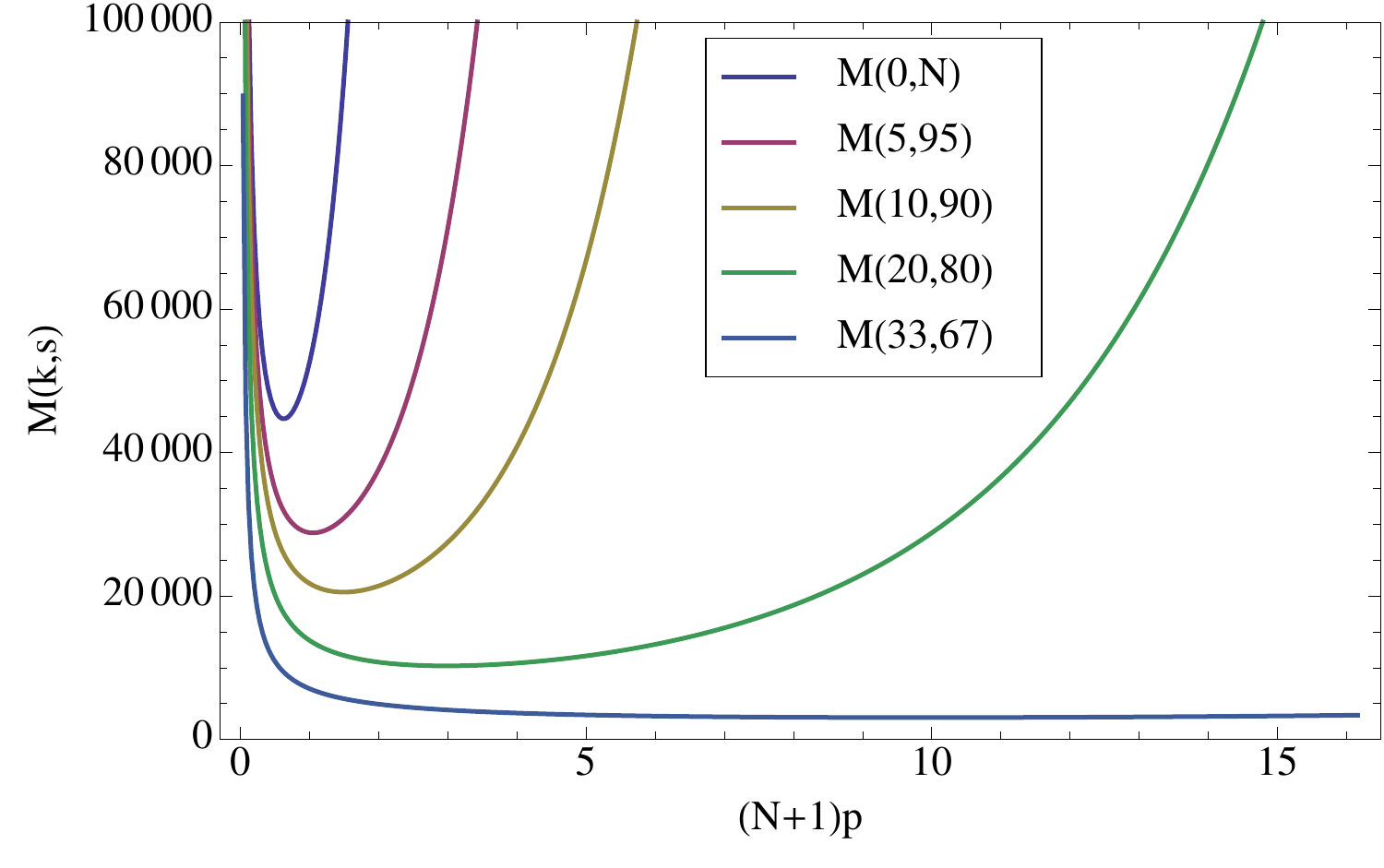}
	\caption{L.h.s.: Mean number of steps required to go from one to the other consensus state as a function of the scaled contrarian rate $(N+1)p$. R.h.s.: Mean number of steps required to go from $X_k$ to $X_s$ as a function of the scaled contrarian rate $(N+1)p$ for $N=100$.}
	\label{fig:CVM.MT.0toN}
\end{figure}

One advantage of using Markov chains as a macro description of the model is that it facilitates the computation of a series of quantities which are more difficult to assess with other techniques.
For the CVM, for instance, we can look at the mean number of steps required to go from one consensus state to the opposite consensus state.
The key to this (and to several other) computations is a matrix called the fundamental matrix \cite{Kemeny1976}:75ff.
For regular chains it is computed by
\begin{equation}
\F = ({\bf I} - (P - W))^{-1}
\label{eq:Fregular}
\end{equation}
where $W$ is the limiting matrix with all rows equal to $\pi$ (note that, $\lim\limits_{n\rightarrow \infty} P^n = W$).
Following \cite{Kemeny1976}, the fundamental matrix can be used to compute another matrix $M$ which contains the mean number of steps between two states, say $i$ and $j$, for any pair of states:
\begin{equation}
M = ({\bf I} - \F + {\bf E} \F_{diag}) D
\label{eq:Mregular}
\end{equation}
where ${\bf E}$ is a matrix with all elements equal to one, $\F_{diag}$ the diagonal fundamental matrix with $(\F_{diag})_{ii} = (\F)_{ii}; (\F_{diag})_{ij} = 0$, and $D$ the diagonal matrix with $(D)_{ii} = 1/\pi_i$.
The mean time from one consensus state to the other is then given by the element $M(0,N)$ which is plotted on the l.h.s. of Fig. \ref{fig:CVM.MT.0toN} for system sizes from $N = 100$ to $N = 500$.
Notice that in Fig. \ref{fig:CVM.MT.0toN} the contrarian rate $p$ is scaled by the size of the macro chain $(N+1)$ in order to compare the different cases.
This accounts for the above-mentioned fact that the >>critical<< parameter value $p^*$ at which a uniform stationary distribution is found depends inversely on the number of agents as $p^* = 1/(N+1)$.
Consequently, in Fig. \ref{fig:CVM.MT.0toN}, the uniform case is represented by $(N+1) p^* = 1$.
%The switching behavior (from one consensus to the other and back) is mainly found for values below that and the behavior approaches the random regime for values larger than one.

%The switching behavior (from one consensus to the other and back) is most frequent for values below $p^*$ and becomes rare as the behavior approaches the random regime for values larger than one.

We observe in the l.h.s. of Fig. \ref{fig:CVM.MT.0toN} that transitions between the two different consensus states are most frequent for a contrarian rate that is slightly below the >>critical<< contrarian rate $p^*$.
Switching becomes very rare as the behavior approaches the random regime ($(N+1) p > 1$), but also for very small $p$.
There is a trade-off between the probability to indeed enter the state of complete consensus and the probability to go away from it and approach to the other extreme.
For the $p$-values where $M(0,N)$ is minimal, both probabilities are relatively high.
As contrarian rate decreases, the probability to reach consensus increases significantly, but a transition to the opposite consensus state is becoming rare.
On the other hand, when $p$ increases slightly, transitions from $k \approx 0$ to $k \approx N$ and back are still rather likely, but in many case the process turns in direction before a complete ordering has been achieved.
This is true also for $p \approx p^*$.
As $p$ increases further, there is a strong decrease in probability to reach consensus altogether (see Fig. \ref{fig:CVM.StationarDist}) and therefore the mean time between the two consensus states increases tremendously.

Finally, the r.h.s. of Fig. \ref{fig:CVM.MT.0toN} shows the same analysis for transitions between states with a strong majority of $\square$-agents to an equally strong majority of $\blacksquare$-agents.
The same qualitative behavior is observed in the sense that switching between strong majorities ($X_0 \leftrightarrow X_N, X_5 \leftrightarrow X_{95}, X_{10} \leftrightarrow X_{90}$) becomes rather unlikely as the contrarian rate increases. 
On the other hand, transitions between moderate majorities of different sign ($X_{20} \leftrightarrow X_{80}, X_{33} \leftrightarrow X_{67}$) occur rather frequently and the contrarian rate at which the mean time between them becomes minimal is larger.

\subsection{Two-Community Model}
\label{cha:5.2Com}

% % REV: intro sentence

We now turn to the two-community graph.
Consider a population composed of two sub-population of size $L$ and $M$ such that $L+M=N$ and assume that individuals within the same sub-population are connected by strong ties whereas only weak ties connect individuals belonging to different communities.
We could think of that in terms of a spatial topology with the paradigmatic example of two villages with intensive interaction among people of the same village and some contact across the villages.
This is similar to the most common interpretation in population genetics where this is called the island model \cite{Wright1943}.
In another reading the model could be related to status homophily \cite{Lazarsfeld1954} accounting for a situation where agents belonging to the same class (social class, race, religious community) interact more intensively than people belonging to different classes.

% % REV: condense, no, reduce, strong reduction

\begin{figure}[h]
	\centering
\includegraphics[width=.5\textwidth]{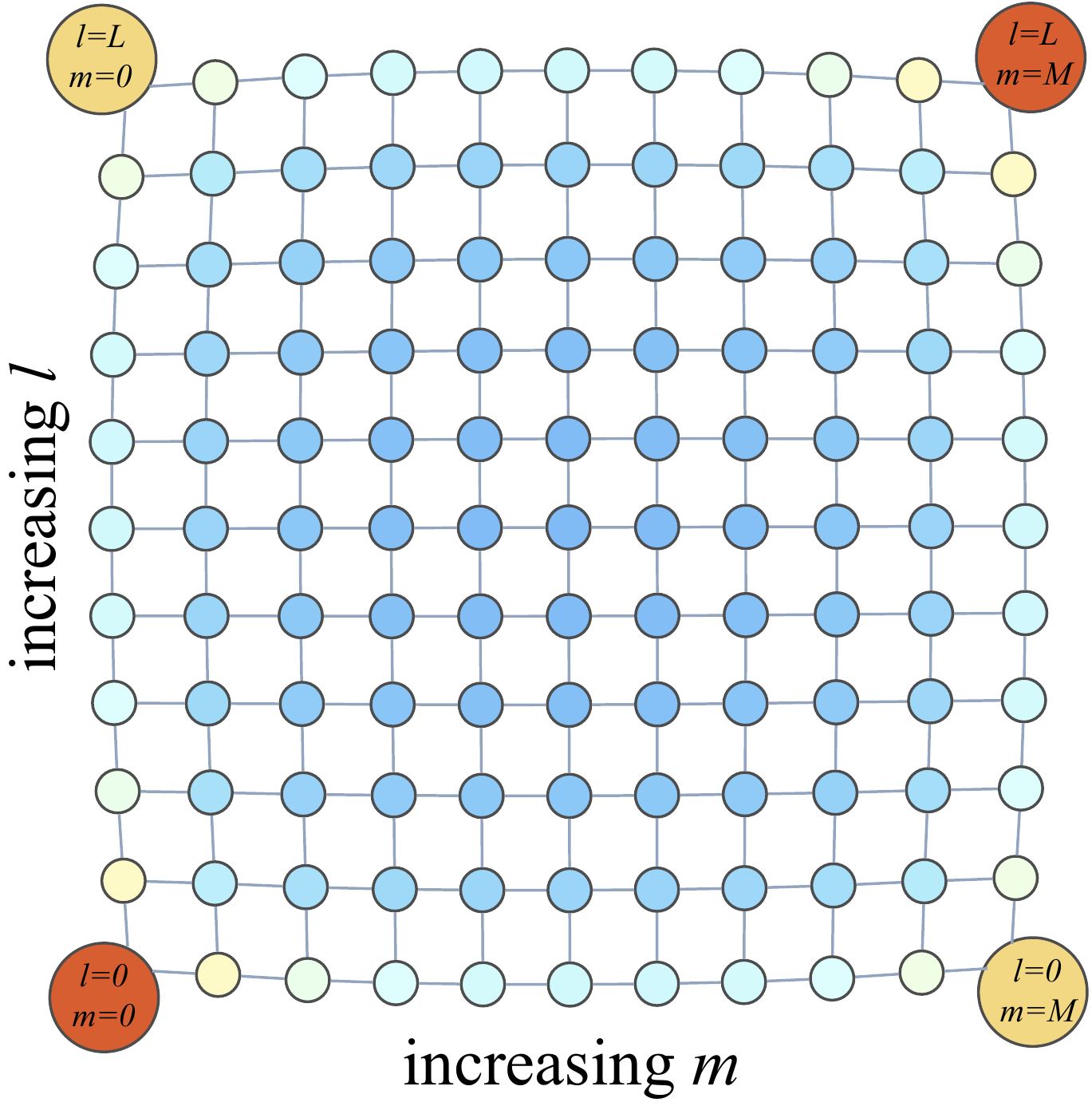}
	\caption{The structure of the CVM meso chain for $L=M=10$. The consensus states $\tilde{X}_{0,0}, \tilde{X}_{M,L}$ as well as the states of  inter-community polarization $\tilde{X}_{0,L}, \tilde{X}_{M,0}$ are highlighted. The stationary distribution is mapped into node colors from blue (low values) to red (high values).}
	\label{fig:Islands_MacroChain20_Fin}
\end{figure}

For the VM, Ref. \cite{Banisch2013acs} shows that a lumpable (that is, Markovian) description for the two-community graph is obtained by a projection of the micro dynamics onto an $(M+1) \times (L+1)$ lattice.
Each lattice point $\tilde{X}_{m,l}$ is associated to the attribute frequencies $m$ and $l$ within the two sub-communities (note that $m+l = k$).
In other words, the model dynamics can be captured without loss of information by a >>mesoscopic<< formulation in terms of attribute frequencies $m$ and $l$ in the two communities.
The state space of the projected model is visualized in Fig. \ref{fig:Islands_MacroChain20_Fin}.

The colors shown in Fig. \ref{fig:Islands_MacroChain20_Fin} represent the stationary distribution of the CVM with a relatively small contrarian probability $p$ and a very weak coupling between the two islands.
The large atoms in the corners of the grid highlight the states that represent configurations of high order.
On the one hand (red-shaded in Fig. \ref{fig:Islands_MacroChain20_Fin}) there are the consensus configuration with all agents in the same state: $\tilde{X}_{L,M}$ and $\tilde{X}_{0,0}$ (in which respectively $k = N$ and $k = 0$).
On the other hand (yellow-shaded), we have the states in which all agents of the same sup-group are aligned, but there is a disagreement across the sub-groups: $\tilde{X}_{0,M}$ and $\tilde{X}_{L,0}$ (for which $k = M$ and respectively $k = L$).
We refer to these states as inter-community polarization.

% % REV: continue shortening

In what follows, we shall refer to the chain shown in Fig. \ref{fig:Islands_MacroChain20_Fin} as \emph{meso chain}  and denote the state space as $\tilde{\X} = (\tilde{X}_{0,0},\ldots,\tilde{X}_{m,l},\ldots,\tilde{X}_{M,L})$.
The notion of >>meso<< in this context accounts for the fact that the process $(\tilde{\X},\tilde{P})$ is indeed in between the micro and the macro level.
Namely, it is a strong reduction compared to the microscopic chain $(\BSigma,\hat{P})$, but the number of states is still considerably larger than the macro system $(\X,P)$ obtained by aggregation over the entire agent population ($h(\x)=k$).
While the full aggregation compatible with homogeneous mixing has led to a random walk on the line with $N+1 = O(N)$ states, the two-community model leads to a random walk on a 2D lattice with $O(N^2)$ states.
Noteworthy, the latter is a proper refinement of the~former.

The transition probabilities of the meso chain are obtained on the basis of Eq. (\ref{eq:PhatVMContrarian}) by substitution of the respective interaction probabilities. % (\ref{eq:gamma.TwoPops}).
%That is  (see \cite{Banisch2013acs}), 
%\begin{eqnarray}
%\gamma = \frac{1}{2 L M+((L-1) L+(M-1) M) r}\\
%\alpha = \frac{r}{2 L M+((L-1) L+(M-1) M) r},
%\label{eq:gamma.TwoPops}
%\end{eqnarray}
Denoting $\omega(i,j) = \gamma$ whenever two agent $i$ and $j$ are in the same community and $\omega(i,j)=\alpha$ whenever they are in different communities, these probabilities are given by (see \cite{Banisch2013acs})
\begin{eqnarray}
\gamma = \frac{1}{2 L M+((L-1) L+(M-1) M) r}\\
\alpha = \frac{r}{2 L M+((L-1) L+(M-1) M) r},
\label{eq:gamma.TwoPops}
\end{eqnarray}
Notice that the ratio between cross-community and intra-community coupling$r = \alpha / \gamma$ is the decisive parameter.
For the CVM on two islands of size $M$ and $L$ the transition probabilities for the transitions leaving the atom $\tilde{X}_{m,l}$ are then given by
\begin{eqnarray}
\begin{array}{l l l}
\tilde{P}(\tilde{X}_{m,l},\tilde{X}_{m+1,l}) & = & (1-p)[\gamma  (m (M-m)) + \alpha  (M-m) l] \\
													&  & + \ p [\gamma (M-m)(M-m-1) + \alpha (L-l)(M-m)]\vspace{6pt}\\
\tilde{P}(\tilde{X}_{m,l},\tilde{X}_{m-1,l}) & = & (1-p)[\gamma  (m (M-m)) + \alpha  m (L-l)]\\ 
													&  & + \ p [\gamma m(m-1) + \alpha l m]\vspace{6pt}\\
\tilde{P}(\tilde{X}_{m,l},\tilde{X}_{m,l+1}) & = & (1-p)[\gamma  (L-l) l + \alpha  (L-l)m] \\ 
													&  & + \ p [\gamma (L-l)(L-l-1) + \alpha (L-l)(M- m)]\vspace{6pt}\\
\tilde{P}(\tilde{X}_{m,l},\tilde{X}_{m,l-1}) & = & (1-p)[\gamma  (L-l) l + \alpha  (M-m)l]\\ 
													&  & + \ p [\gamma l(l-1) + \alpha l m].
\end{array}
\label{eq:TransProbs.T2.TwoPopCVM}
\end{eqnarray}

%\begin{eqnarray}
%\gamma = \frac{1}{2 L M+((L-1) L+(M-1) M) r}\\
%\alpha = \frac{r}{2 L M+((L-1) L+(M-1) M) r},
%\label{eq:gamma.TwoPops}
%\end{eqnarray}

\subsection{Stationary Dynamics on the Two-Community Graph}
\label{cha:5.Pi.2Com}

% % REV: shortening + slight clarifications

As described in Sec. \ref{cha:5.HMStatDyn}, the stationary distribution $\pi$ of a Markov chain %with transition matrix $P$ is the probability vector $\pi$ that satisfies $\pi P = \pi$ the computation of which 
requires the computation of the left eigenvector of $P$.
In what follows we study the two-community model with $M = L = 50$, the Markov projection of which results in a Markov chain of size $(M+1)(L+1) = 2601$.
For a matrix of size $2601 \times 2601$ the (numerical) solution of the corresponding eigenvalue problem is still possible, but increasing the number of agents (that is, $M$ and $L$) will soon lead to matrix sizes for which the solution for eigenvalues and vectors is rather costly.

\begin{figure}[htp]
	\centering
\includegraphics[width=1.0\textwidth]{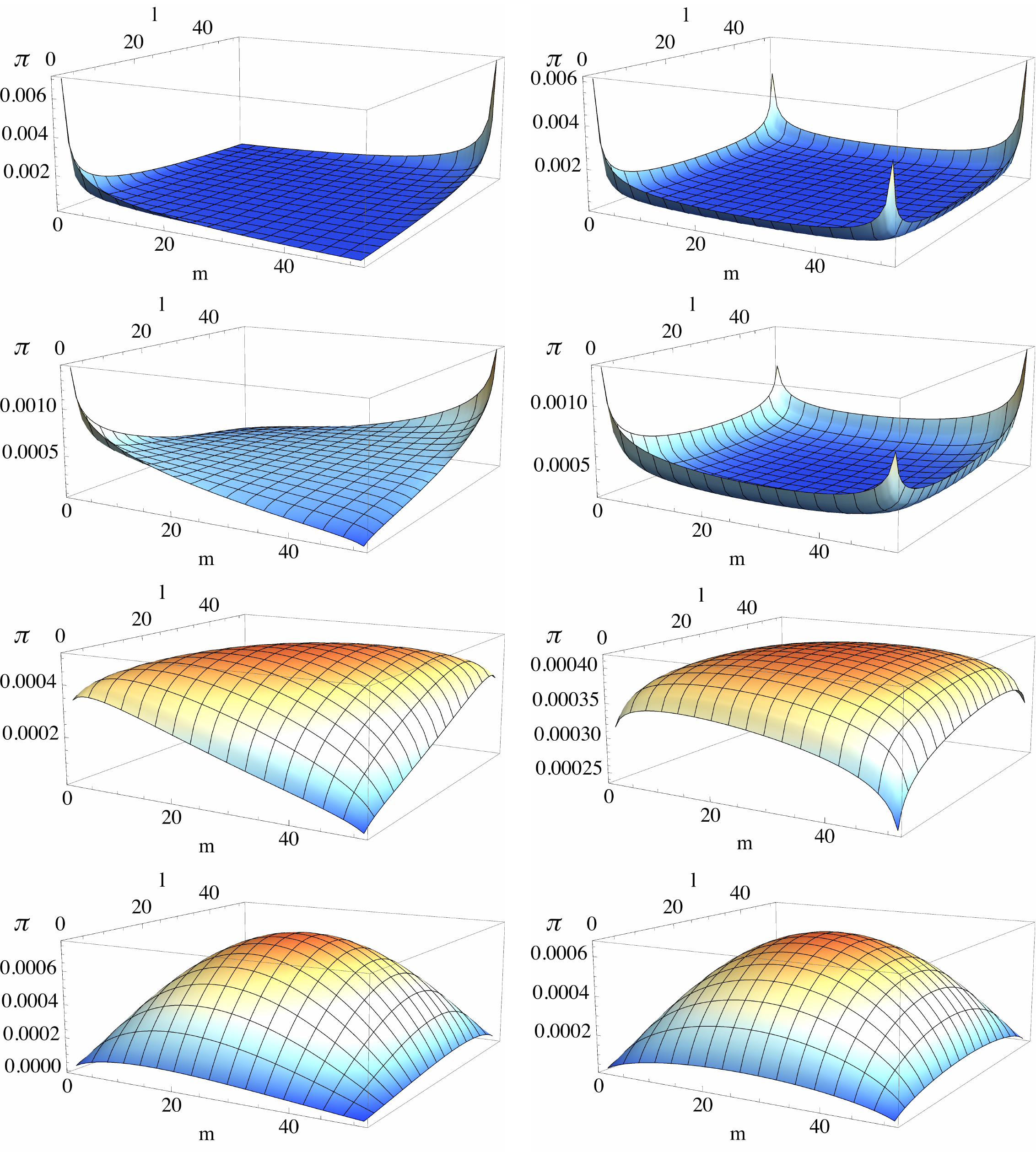}
	\caption{Stationary distribution for different $p$ and $r$ for a system of $M = L = 50$. The column on the l.h.s. is for a moderate coupling $r = 1/100$ and the four plots on the r.h.s. are for a weak coupling $r = 1/1000$. From top to bottom the contrarian rates are $p = 0.01,0.015,0.02,0.03$. The stationary probability for the consensus states ($m = l = 0$ and $m = l = 50$) increases with decreasing $p$. The stationary probability for the states of partial order ($m = M, l = 0$ and $m = 0, l = L$) increases as the coupling between the island $r$ decreases. This topological effect is undermined by an increasing contrarian rate $p$.}
	\label{fig:Islands.CVM.Pi}
\end{figure}

There are two parameters that decide about the dynamical behavior of the CVM on the two-community graph: (i.) the contrarian rate $p$, and (ii.) the coupling between the two islands captured by $r = \alpha/\gamma$.
To obtain a complete picture of the model dynamics, the stationary distribution has been computed for various different values $p$ and $r$ which is shown in Fig.~\ref{fig:Islands.CVM.Pi}.
% % REV:
In accordance with Fig. \ref{fig:Islands_MacroChain20_Fin}, the ordered states (complete consensus and inter-community polarization) correspond to the corners in these plots. 
From the top to the bottom, $p$ is increased from $p = 0.01$, $p = 0.015$, $p = 0.02$ to $p = 0.03$.
The plots in the left-hand column show the result for a moderate coupling between the two island with $r = 1/100$. 
A reduced coupling of $r = 1/1000$ is shown in the plots in the right-hand column.

The comparison of the left- and the right-hand side of Fig. \ref{fig:Islands.CVM.Pi} shows that the stationary probability for states of inter-community polarization, as well as the states close to them, increases with a decreasing coupling between the communities.
That is, the configurations with intra-community consensus, but disagreement across the communities become more and more probable.
This is very obvious for the plots with a small contrarian rate $p = 0.01$ and $p=0.015$ where the probability to observe the states $\tilde{X}_{0,50}$ or $\tilde{X}_{50,0}$ becomes very high when decreasing the coupling to $r = 1/1000$.
In fact, all configurations in which consensus is formed in at least one of the communities are rather likely (values along the border of the surface) whereas disordered configurations are rare.
This is in direct analogy to the pure VM ($p=0$) where states of inter-community polarization become meta-stable states in which the process is >>trapped<< for some time (see \cite{Banisch2013acs}).
However, the significant difference between a moderate ($r = 1/100$) and a very weak ($r = 1/1000$) coupling diminishes as the contrarian rate becomes larger.
This second trend observed in Fig. \ref{fig:Islands.CVM.Pi} is in agreement with what happens in the homogeneous mixing case as the contrarian rate $p$ increases: the probability to observe consensus configurations with all agents in equal state becomes more and more unlikely and it is more and more likely to observe disordered agent configurations all together.
In fact, a further increase of the contrarian rate will lead to a behavior that is essentially random and insensitive to topological constraints since the consensus formation within the communities is frequently perturbed by random events.

% % REV: relation to previous work

\enlargethispage{\baselineskip}
It would be very interesting to perform the two-community analysis for other opinion models and different kinds of nonconformity behavior.
For the Galam and the Sznajd model with contrarians (see \cite{Galam2004} and \cite{Nyczka2012}) we hypothesize a more interesting distribution in which the four peaks are located at mixed minority-majority configurations.
Furthermore, it would be interesting to analyze the effect of independent agents as studied in the $q$--voter model in \cite{Nyczka2012}.
Namely, on the complete graph (for $q > 5$) a third peak emerges centered at the fifty-fifty profiles and it is not clear how this effect translates on the two--community graph.

\section{Network Dynamics from the Macro Perspective}
\label{cha:5.NetworkDynamics}

\subsection{Network Influence on the Stationary Dynamics}
\label{cha:5.Pi.Networks}

Let us consider, in a numerical experiment, the effect of different interaction topologies $\omega$ on the stationary dynamics of the resulting macro process.
For this purpose, we define the stationary macro measure $\pi$ as
\begin{equation}
\pi_k = \sum\limits_{\x \in X_k}^{} \hat{\pi}_{\x}.
\label{eq:Pi.MacroNets}
\end{equation}
In other words, the elements $\pi_k$ of the stationary vector are determined by counting the frequency with which the model is in the respective set of micro states $\x \in X_k$.
Notice that on the basis of a stationary micro chain, it is always possible to construct an approximate macro chain -- an aggregation -- the stationary vector of which satisfies Eq. (\ref{eq:Pi.MacroNets}) (see \cite{Kemeny1976}:140 and \cite{Buchholz1994}:61--63).
%This will be discussed below.

\begin{figure}[hbtp]
\centering
\includegraphics[width=0.75\linewidth]{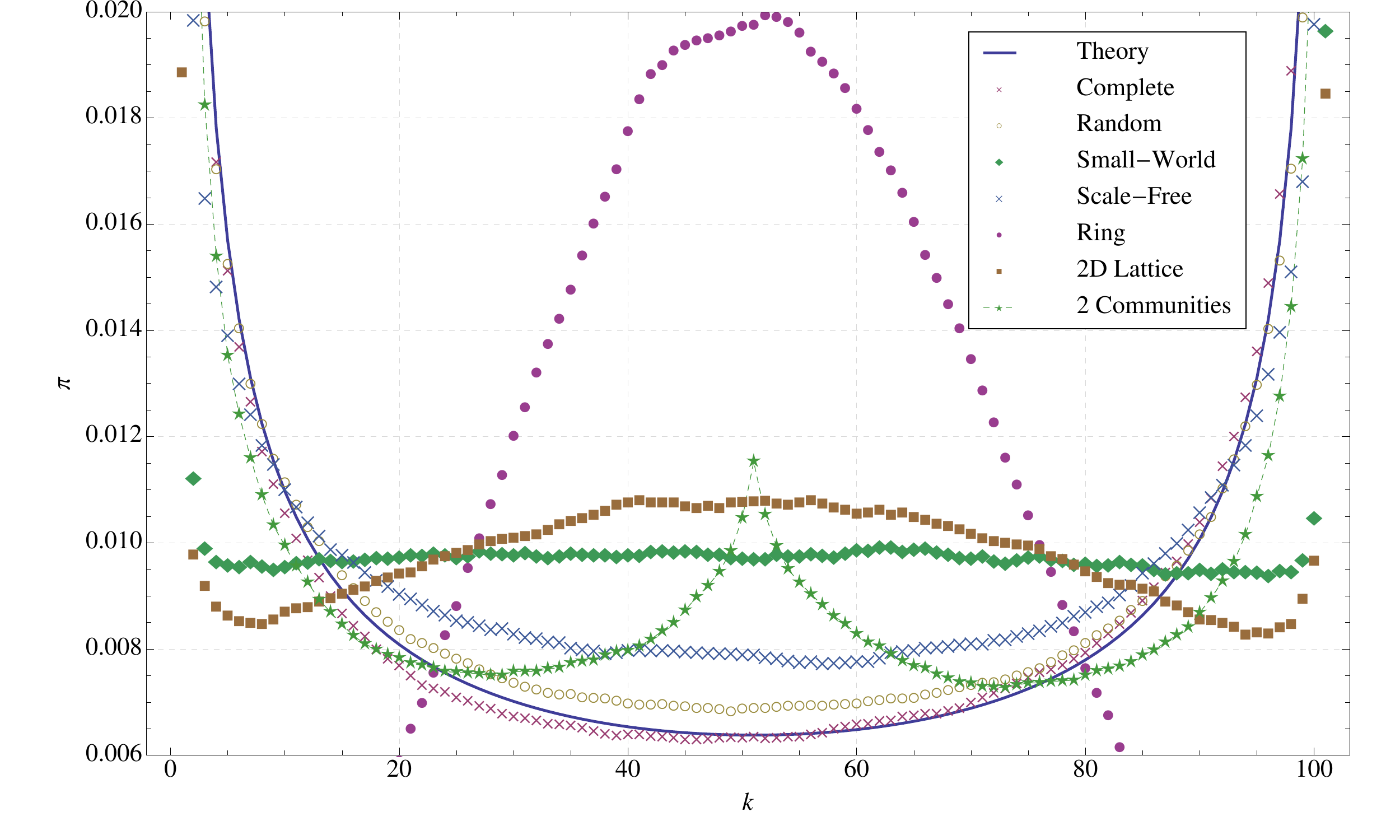}
\caption{Stationary statistics for the CVM on different topologies. Due to effects of local alignment, the stationary behavior of the small-world network, the ring and the lattice as well as the two-community topology differs greatly from the well-mixed situations.}
\label{fig:p005}
\end{figure}

To compute the $\pi_k$, a series of simulations has been performed in which the CVM with $N = 100$ is run on different paradigmatic agent networks.
To capture the model in stationarity, the model is iterated for several thousands of steps first and the statistics of this >>burn-in<< phase are not considered in the computation of $\pi_k$.
(In this exploratory analysis with 100 agents a >>burn-in<< period of 20000 steps has been used.)
The result is shown in Fig. \ref{fig:p005} for the case of a small contrarian rate $p =0.005$.

We observe in Fig. \ref{fig:p005} that some interaction topologies give rise to strong deviations from the theoretical result derived for homogeneous mixing (solid, blue).
In general, there is an increase in the probability to observe balanced configurations and the case of complete consensus tends to become less likely.
However, the results obtained for the random graph are indeed very similar to the theoretical prediction and also the scale-free topology leads to stationary statistics that, in qualitative terms, correspond to the mean-field case.
On the other hand, we observe a strong >>modulation<< of the stationary statistics by networks that tend to foster the emergence of >>local alignment and global polarization<<.
By local alignment and the dynamics that lead to it, we refer to situations in which different clusters of agents approach independently a certain local consensus which is in general different from agent cluster to agent cluster.
From the global perspective the entire population appears to be far from complete consensus and the probability to observe the respective intermediate macro states is increased.
These effects are observed for the small-world network, the two-community graph as well as for the lattice, and it is strongest for the ring where the probability of complete consensus is practically zero.

\subsection{The Two-Community Case}

For the two-community graph with a peak around $k = N/2$ the interpretation of the result is particularly straightforward.
Local alignment, in this case, refers to inter-community polarization -- the situation in which a different consensus has emerged in the two communities.
If the size of the communities is $N/2$, as in the example we study, the polarization configurations give rise to an macro observation $m+l = k = N/2$ since one half of the population (organized in one community) agrees on $\square$ and the other half (that is, the other community) on $\blacksquare$.
%We will come back to this in Sec. \ref{cha:5.MesoMacro} where ...

\begin{figure}[htp]
	\centering
	\begin{tabular}{l c r}
\includegraphics[width=.33\textwidth]{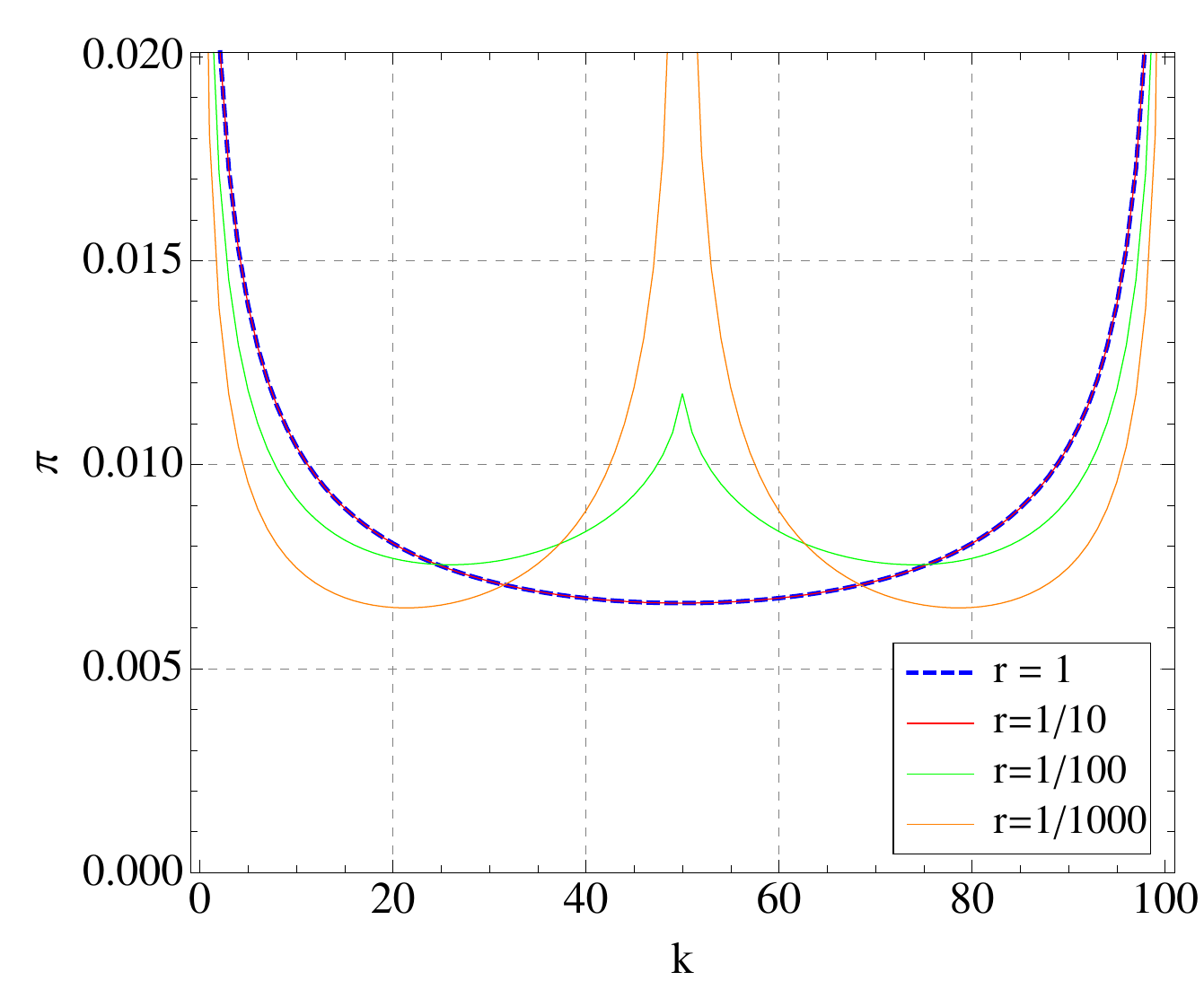}&\hspace{-12pt}\includegraphics[width=.33\textwidth]{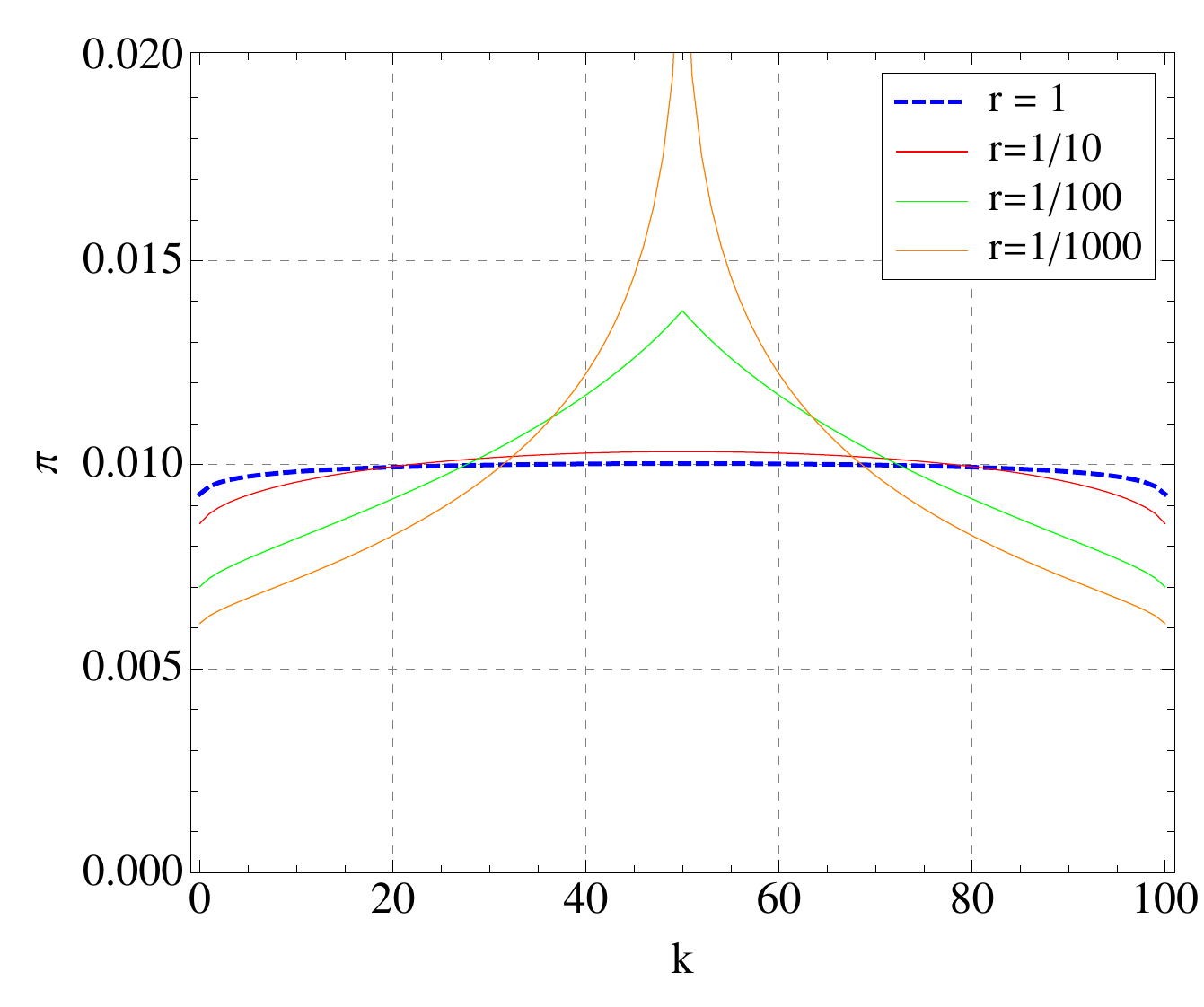}&\hspace{-12pt}\includegraphics[width=.33\textwidth]{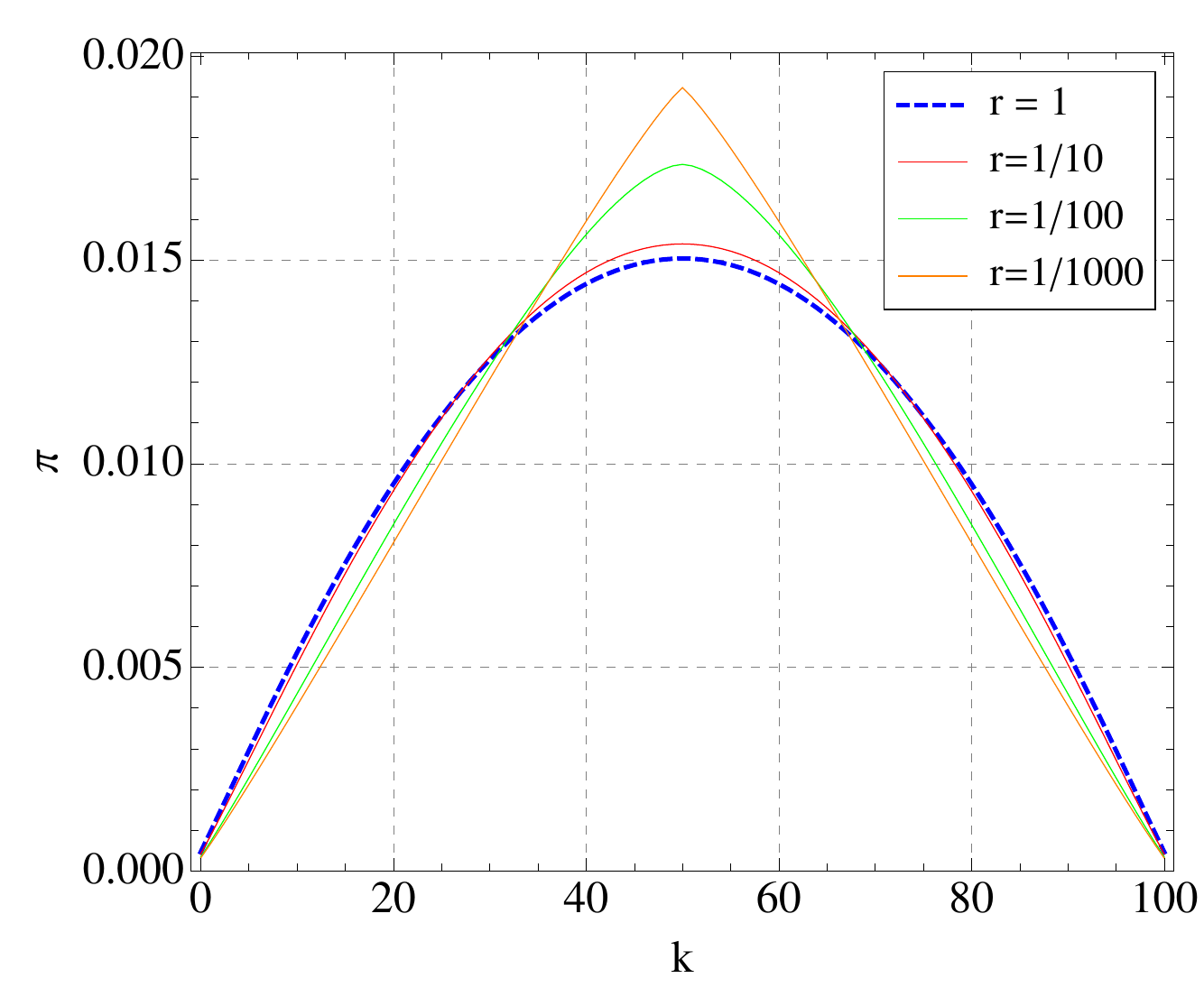}
	\end{tabular}
	\caption{The stationary distribution from the macro perspective for different $r = \alpha/\gamma = 1, 1/10, 1/100, 1/1000$. From left to right $p = 0.005, 0.01,0.02$.}
	\label{fig:CVM_PiComp}
\end{figure}

The two-community CVM is particularly interesting because we can compute the exact stationary vector by analyzing the respective meso chain $(\tilde{\X},\tilde{P})$ obtained via strong lumpability (see Secs. \ref{cha:5.2Com} and \ref{cha:5.Pi.2Com}).
%This has been done in Sec. \ref{cha:5.Pi.2Com}.
We first compute the stationary distribution of the meso chain assigning the respective limiting probability $\tilde{\pi}_{m,l}$ to each state $\tilde{X}_{m,l}$.
In that case, Eq. (\ref{eq:Pi.MacroNets}) reads
\begin{equation}
\pi_k = \sum\limits_{m+l = k}^{} \tilde{\pi}_{m,l}.
\end{equation}
That is, $\pi_k$ associated to the macro state $X_k$ is obtained by summing up the respective $\tilde{\pi}_{m,l}$ with $m+l=k$.
This is shown in Fig. \ref{fig:CVM_PiComp} for different contrarian rates $p$ and different couplings between the two sub-graphs.

It becomes clear that the probability to observe a fifty-fifty situation ($k \approx N/2$) generally increases, the weaker the coupling between the communities.
The analysis of the meso level stationary distribution shown in Fig. \ref{fig:Islands.CVM.Pi} makes clear that this is due to an increased probability for the configurations with intra-community consensus and inter-community polarization ($\tilde{X}_{N/2,0},\tilde{X}_{0,N/2}$) which contribute to that probability.
(Notice that the community sizes have a direct effect onto the macro level statistics and that in general the states with $k = L$ and $k = M$ will be observed more frequently when the coupling is weak.)
In general, we can also observe that the influence of the different topological choices onto the macro behavior (captured here in terms of $\pi_k$) decreases with an increasing contrarian rate $p$.
As explained in Sec. \ref{cha:5.Pi.2Com}, the more contrarian behavior is allowed by the parameter setting, the more random becomes the entire process which undermines the effects of local alignment and, consequently, of interaction topology.
This can be taken is an indication that the homogeneous mixing solution (here represented by $r=1$) might approximate well the model behavior with a relatively high contrarian rate because the entire setting is characterized more and more by random state flips.
It will be less accurate for a small contrarian rate where local alignment becomes more characteristic.

% % REV: Clearer title?

%\section{The Two-Community Model as an Analytical Framework}
\section{Quantification of Non-Markovianity in the Two-Community Model}
\label{cha:5.MicroMesoMacro}

\subsection{From Micro to Meso, and from Meso to Macro}

% % REV: make motivation clearer

The previous section has shown that heterogeneous interaction structures can have a strong impact on the model behavior.
From the lumpability point of view, but also from the point of view of observation, a macro process obtained by global aggregation over the agent attributes neglects important information about the microscopic details.
The remainder of this paper is an attempt to better understand this loss of information and the macro-level effects this leads to.

Even though the questions addressed in this section may be not directly relevant for an interpretation in terms of opinion dynamics, an improved understanding of the dynamical effects introduced by the way an agent system is observed is of great relevance for ABMs more generally.
The simple rules of the CVM along with the controllable two-community topology make this scenario well-suited for a first step to analyze the effects at an aggregate level introduced by aggregation without sensitivity to micro- or mesoscopic structures.

%In most binary opinion models (as well as in other binary models) the observable of interest is the opinion frequency obtained by global aggregation over the entire population, as described in Sec. \ref{sec:2.FullAgg}.
%In general, this global aggregation provides us with an exact macro description only in the absence of inhomogeneities.

%is the homogeneous mixing case which has been addressed in Sec. \ref{cha:5.HMand2Com}.

%We have also discussed in Sec. \ref{cha:5.HMand2Com} the two-community graph for which a meso level chain with $O(N^2)$ states provides an loss-less coarse-graining.
%As the two-community coarse-graining ($\tilde{\X}$) is a proper refinement of the full aggregation ($\X$), and for it is still tractable, we may use this topology as a test case to address questions that concern the relation between the two coarse-grainings.
%Having available an aggregated mean-field-like description of the process $(\X,P)$ on the one hand, and a bigger Markov chain that describes exactly the model dynamics on an interaction topology with a small amount of inhomogeneity $(\tilde{\X},\tilde{P})$ on the other, we can also analytically assess how the topology effects the macro dynamics and in which regard the network dynamics deviate from the mean-field behavior.

\begin{figure}[htp]
	\centering
\includegraphics[width=0.95\textwidth]{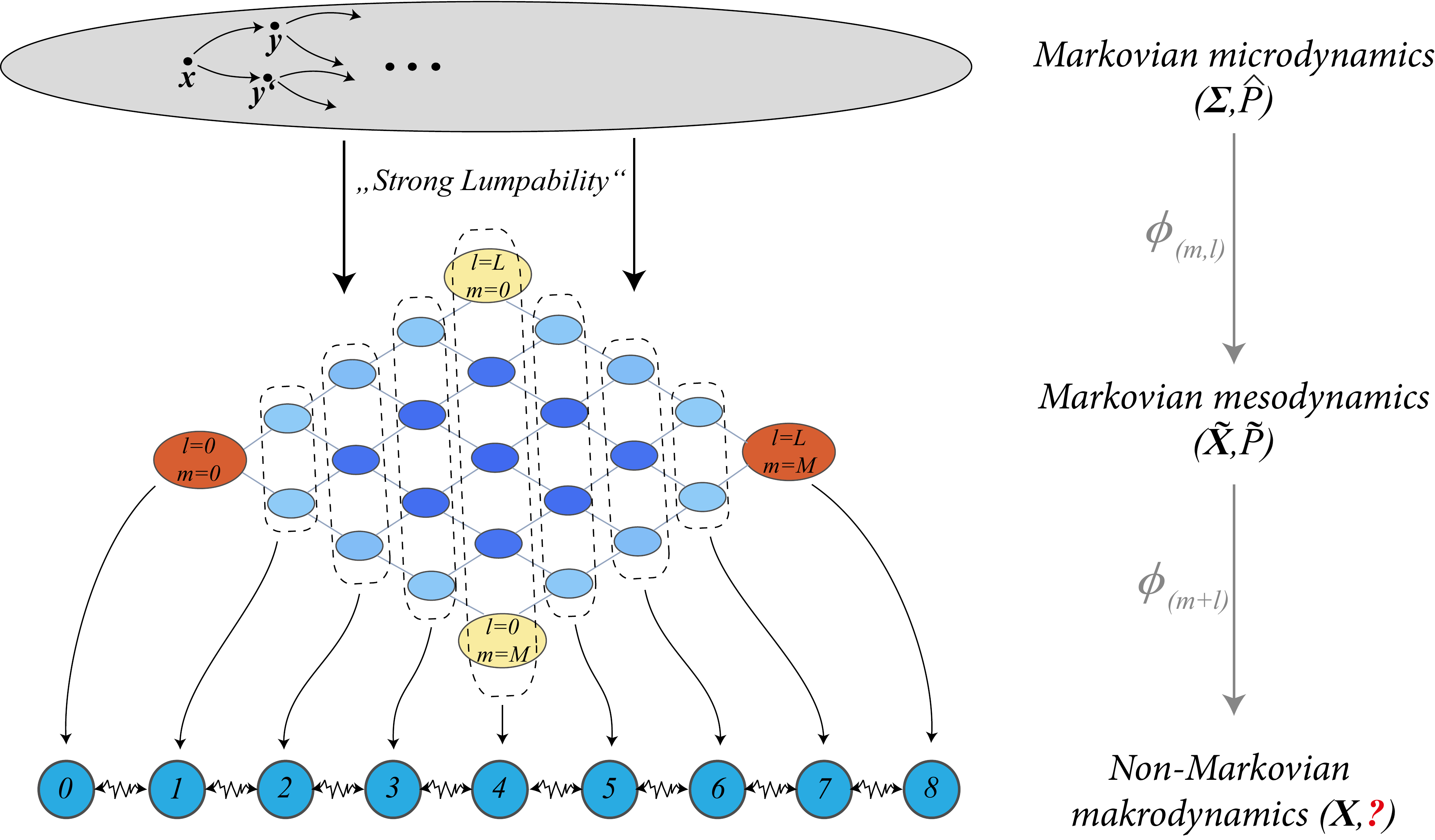}
	\caption{From micro to meso, and from meso to macro.}
	\label{fig:Islands.Projection}
\end{figure}

The general idea is illustrated in Fig. \ref{fig:Islands.Projection}.
Consider the CVM on the two-community graph and the associated micro-level process $(\BSigma,\Phat)$.
The two-community micro chain $(\BSigma,\Phat)$ is (strongly) lumpable with respect to the partition $\tilde{\X}$.
This gives rise to what we have called the meso-level process $(\tilde{\X},\tilde{P})$ in Sec. \ref{cha:5.2Com}.
The meso chain gives us a complete understanding of the (micro) behavior of the CVM on two coupled communities, because the coarse-graining via strong lumpability is compatible with the exact symmetries of the micro process.
That is, no information is lost by a formulation of the dynamics in terms of the frequencies $m$ and $l$ in the two communities.
However, the process (the micro as well as the meso chain) is not lumpable with respect to the macro level of full aggregation (partition $\X$) which formulates the dynamics in terms of the opinion frequency in the entire population ($k = m+l$).
Therefore, if we wish to observe the process at the global level, which is often the case in binary opinion models (as well as in other binary models), we must live with the fact that the resulting macro process on $\X$ is no longer a Markov chain.
In other words, more complex temporal correlations (memory effects) emerge at the macro level.

As illustrated in Fig. \ref{fig:Islands.Projection}, here we project onto the level of full aggregation despite the fact that Markovianity is lost, in order to understand (i.) the reasons for which lumpability is violated and (ii.) the dynamical effects that this introduces at the macro level.
That is, all meso states $\tilde{X}_{m,l}$ with the same global opinion frequency $k = m + l$ are projected into the same macro state $X_k$.
The fact that we have an explicit understanding of the meso chain facilitates an explicit analysis of the transition from micro to meso to macro.

\subsection{Why Lumpability Fails}

\begin{figure}[htp]
	\centering
\includegraphics[width=.8\textwidth]{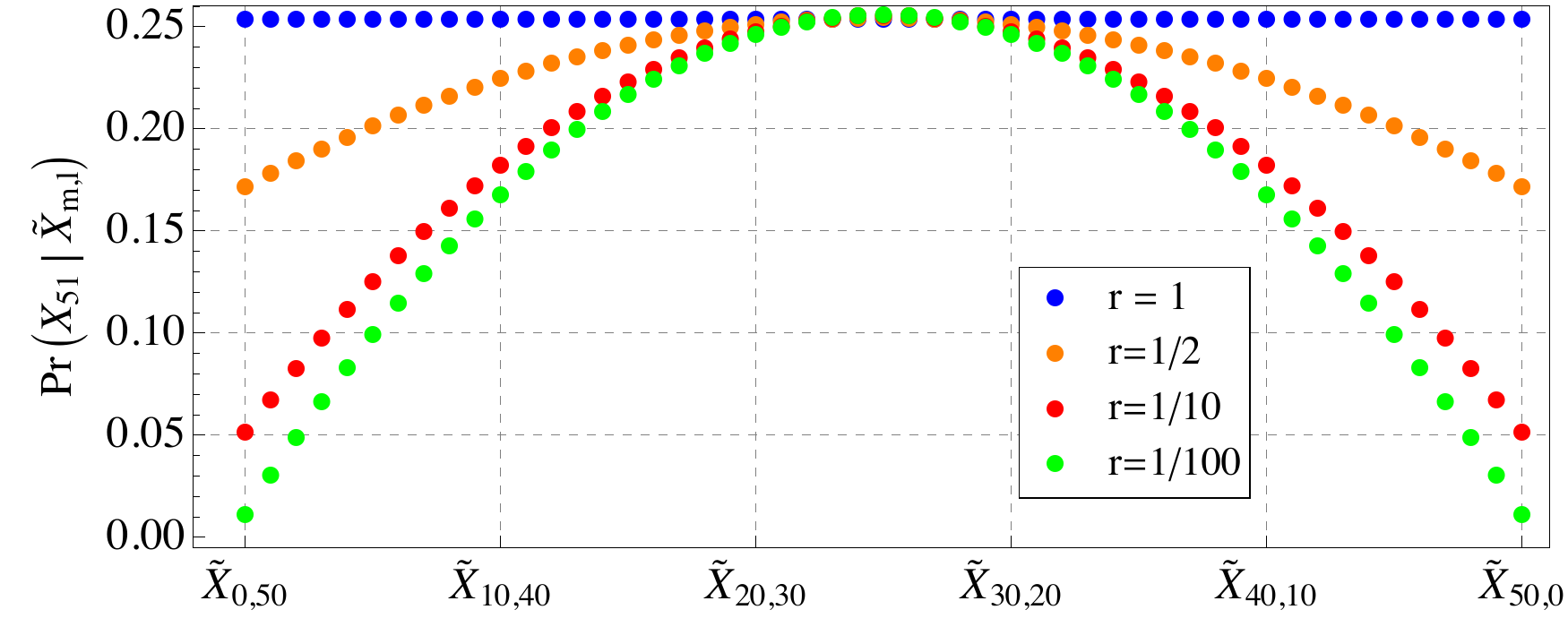}
	\caption{The island topology leads to inhomogeneous transition probabilities and is therefore not (strongly) lumpable. Here the example of a transition from $\tilde{X}_{m,l} \in X_{50}$ to $X_{51}$ in a system with $M = L = 50$ is shown.}
	\label{fig:CVM_InhomP_p001}
\end{figure}

Let us first inspect the reasons for which the meso chain $(\tilde{\X},\tilde{P})$ is not lumpable with respect to the macro partition $\X$.
By the lumpability theorem (Thm. 6.3.2 in \cite{Kemeny1976}), it is clear that the non-lumpability of the meso chain with respect to $\X$ comes by the fact that the probabilities $Pr(X_k|\tilde{X}_{m,l})$ are not equal for all meso states $\tilde{X}_{m,l} \in X_{(m+l)}$ in the same macro set.
As an example, we consider the transition rates from the single $\tilde{X}_{m,l} \in X_{50}$ to the macro set $X_{51}$ in a system with $M = L = 50$.
One by one, the conjoint probability from $\tilde{X}_{0,50},\tilde{X}_{1,49},\tilde{X}_{2,48}, \ldots, \tilde{X}_{50,0}$ to the sets $\tilde{X}_{m,l} \in X_{51}$ is shown in Fig. \ref{fig:CVM_InhomP_p001} for various $r$ and a small contrarian rate $p = 0.01$.

We first notice that the transition rates $\tilde{P}(\tilde{X}_{m,l},X_{(m+l+1)})$ are uniform when the coupling within is equal to the coupling across communities, that is for $\alpha = \gamma$ and $r=1$.
Obviously, this is the case of homogeneous mixing and the uniformity of the $\tilde{P}(\tilde{X}_{m,l},X_{(m+l+1)})$ is precisely the lumpability condition spelled out in Thm. 6.3.2 of \cite{Kemeny1976}.

In general, the $\tilde{P}(\tilde{X}_{m,l},X_{(m+l+1)})$ are no longer equal for all $m$ and $l$ with $m+l = k$ when heterogeneity is introduced in form of a different coupling within and across communities, i.e., $\alpha \neq \gamma$.
This explains the non-lumpability of the two-community model with respect to $\X$.
As the weak ties across communities becomes weaker such that the ratio $r$ between strong and weak ties decreases, the transition rates become inhomogeneous, the main effect being a strong decrease of $\tilde{P}(\tilde{X}_{m,l},X_{(m+l+1)})$ for the atoms close to polarization ($m = M, l=0$ and $m = 0,l = L$).
This decrease in transition probability, in turn, explains the increased stationary probability of the states $\tilde{X}_{50,0}$ and $\tilde{X}_{0,50}$ observed in Sec. \ref{cha:5.Pi.2Com}, because once entered there is a relatively small probability to leave them so that the process is likely to >>wait<< in these states for quite some time.

Notice that there is only the small difference in transition rates between $r = 1/10$ and $r = 1/100$ (the difference to $r = 1/1000$ is even smaller!).
On the one hand, this is somewhat surprising, as from the dynamical point of view $r = 1/10$ is much more related to the homogeneous mixing case ($r = 1$) than to the situation with $r = 1/100$ (cf. Fig. \ref{fig:CVM_PiComp}).
On the other hand, the probability to leave a polarized state ($\tilde{X}_{0,50},\tilde{X}_{50,0}$) decreases significantly with every decrease in $r$ and therefore the waiting times for these states grow tremendously.
Notice, however, that in the limit of $r \rightarrow 0$, the probability of leaving a polarized state converges to $p$ (with $\tilde{P}(\tilde{X}_{M,0},X_{(M+1)}) = \tilde{P}(\tilde{X}_{M,0},X_{(M-1)}) = p/2$).
Therefore a strong difference between a weak (e.g., $r = 1/100$) and a very weak coupling ($r = 1/1000$) in form of an increased stationary probability of polarization can be expected only if also the contrarian rate $p$ is small.
Likewise, as already observed in Sec. \ref{cha:5.Pi.2Com}, a large contrarian rate can completely undermine effects of polarization altogether.

% % REV: comment on weak lumpability

Finally, notice that the two-community CVM does also not satisfy the conditions of weak lumpability which refers to the fact that a Markov chain might be lumpable only for particular starting vectors \cite[]{Burke1958,Kemeny1976,Ledoux1994}.
Based on the fact that if a chain is weakly lumpable with respect to some distribution, it must be lumpable with respect to the stationary distribution, it is easy to construct a counter example which shows that the conditions for weak lumpability are violated (see \cite{Banisch2014phd}, pp. 112-117 for details).

\subsection{Measuring (Non)-Markovianity}
\label{cha:5.MarkovMeasure}

%The previous section shows that  the macro process associated to the CVM on two coupled communities is non-Markovian.
%The next logical step is to quantify in some way the deviations from Markovianity.
Having shown that the macro process associated to the CVM on two coupled communities is non-Markovian, the next logical step is to quantify in some way the deviations from Markovianity.
The framework of information theory -- relative entropy and mutual information in particular -- has been shown to be quite useful for this purpose \cite[among others]{Chazottes1998,Vilela2002,Goernerup2008,Ball2010,James2011,Pfante2013}.

Let us, to simplify the writing, denote as $[\ldots,k_{t-2},k_{t-1},k_t,k_{t+1},k_{t+2,\ldots}]$ a sequence of macro states $\ldots \rightarrow X_{k_{t-2}} \rightarrow  X_{k_{t-1}}  \rightarrow  X_{k_{t}}  \rightarrow  X_{k_{t+1}}  \rightarrow  X_{k_{t+2}}  \rightarrow \ldots$.
Likewise, let us denote as $[k_{t-m},\ldots,k_t]$ a finite sequence of $m$ macro states and refer to this as block or cylinder of length $m$.
Then, the block entropy associated to cylinders of length $m$ is defined by
\begin{equation}
H_m = \sum\limits_{[k_{t-m},\ldots,k_t] \in \GG_m}^{} \mu([k_{t-m},\ldots,k_t]) \log \mu([k_{t-m},\ldots,k_t])
\label{eq:BlockEntropy}
\end{equation}
where $\mu([k_{t-m},\ldots,k_t])$ denotes the probability to observe the respective cylinder $[k_{t-m},\ldots,k_t]$.
Notice that for $m > 1$ there exist in general >>forbidden<< sequences with $\mu([k_{t-m},\ldots,k_t]) = 0$, a fact that is usually formalized in terms of a grammar $\GG_m \subseteq \X^m$ by defining $\GG_m := \{[k_{t-m},\ldots,k_t] : \mu([k_{t-m},\ldots,k_t]) > 0\}$.
In our case of single-step dynamics, all sequences containing subsequent elements with $|k_t - k_{t-1}| > 1$ are >>forbidden<< because only $X_k, X_{k-1}$ and $X_{k+1}$ can be reached from $X_k$ in one step.

It is well-known \cite[]{Chazottes1998,Vilela2002,James2011} that the slope of the block entropy $\Delta H_m = H_m - H_{m-1}$ converges to a fixed value called entropy rate (usually denoted as $h(\mu)$) and that this fact can be used to estimate the memory range of the process.
Namely, following \cite{Chazottes1998,Vilela2002}, the range of the process is given, at least in an approximative sense, by the $m$ at which $\Delta H_m$ reaches a constant value, that is, $\Delta H_m - \Delta H_{m+1} \approx 0$.
It is clear then that for a Markovian process this point must be reached at $m = 2$ such that
\begin{equation}
\Delta H_2 - \Delta H_3 = 0
\label{eq:Derivation.I1}
\end{equation}
and more generally
\begin{equation}
\Delta H_2 - \Delta H_m = 0.
\label{eq:Derivation.In}
\end{equation}

Notice that Eq. (\ref{eq:Derivation.In}) is precisely the >>Markov property measure<< proposed in \cite{Goernerup2008}, pp.6-8, to identify projections of a process onto a smaller state space (a partition of the original process) which lead to Markovian dynamics.
Noteworthy, the starting point of \cite{Goernerup2008} is the expected mutual information $\langle I\rangle$ between pasts and the future state.
Namely, how much information about the next symbol ($[k_{t+1}]$) is on average over all symbols contained in the sequence of symbols ($[\ldots,k_{t-2},k_{t-1}]$) before the current symbol ($[k_t]$).
They show that the expected past future mutual information can be expressed in terms of the slopes of the block entropy as
\begin{equation}
\langle I\rangle = \Delta H_2 - \Delta H_{\infty}
\label{eq:I}
\end{equation}
and likewise 
\begin{equation}
\langle I_n\rangle = \Delta H_2 - \Delta H_{2+n}
\label{eq:In}
\end{equation}
if finite histories of length $n$ ($[k_{t-n}\ldots,k_{t-2},k_{t-1}]$) are considered.\footnote{In real computations, one always has to restrict to finite histories. \cite{Goernerup2008} compute $\langle I_2\rangle = \Delta H_2 - \Delta H_{4}$ which means that they consider cylinders up to length four $[k_{t-2},k_{t-1},k_{t},k_{t+1}]$ in their Markovianity test.}
Notice that in their notation $n$ accounts for the ranges beyond the Markov range of two ($m = n+2$ in Eq. \ref{eq:Derivation.In}).
We will follow this notation here and compute $\langle I_1\rangle = \Delta H_2 - \Delta H_{3}$ and $\langle I_2\rangle = \Delta H_2 - \Delta H_{4}$, the latter being used by \cite{Goernerup2008}.

The advantage of the two-community CVM as a framework to link between a micro and a macro level of description via an intermediate meso level description is that we are able to \emph{compute} the Markovianity measures $\langle I_1\rangle$ and $\langle I_2\rangle$ instead of performing an extensive series of numerical simulations.
Namely, it is possible to compute the $\mu([k_{t-1},k_{t},k_{t+1}])$ and respectively the $\mu([k_{t-2},k_{t-1},k_{t},k_{t+1}])$ on the basis of the meso chain $(\tilde{\X},\tilde{P})$ %which is a loss-less description of the original microscopic system 
(see Fig. \ref{fig:Islands.Projection}).

\begin{figure}[ht]
	\centering
\includegraphics[width=.34\textwidth]{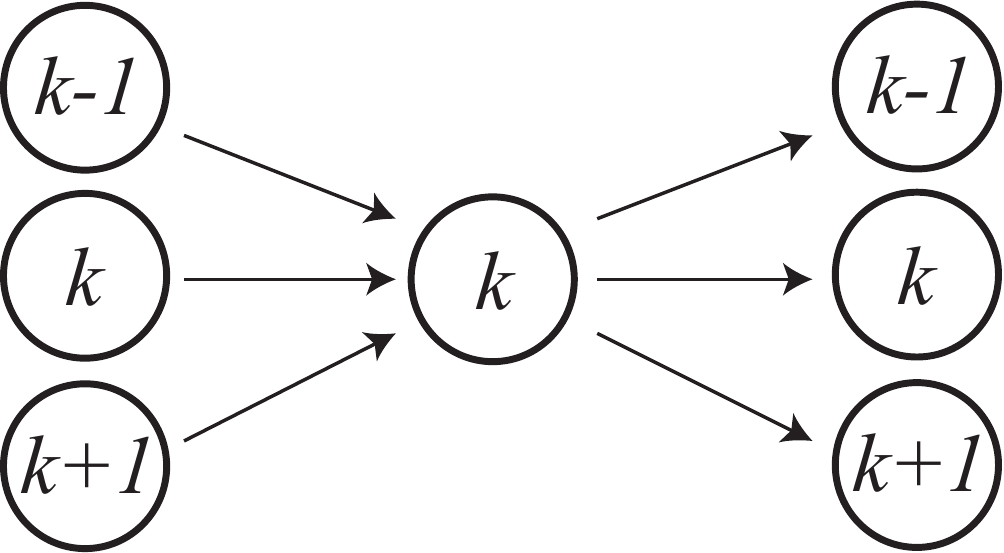}
	\caption{Possible paths of length 3 through $X_k$.}
	\label{fig:Macro3Cylinder}
\end{figure}

\begin{figure}[ht]
	\centering
\begin{tabular}{l r}
\includegraphics[width=.5\textwidth]{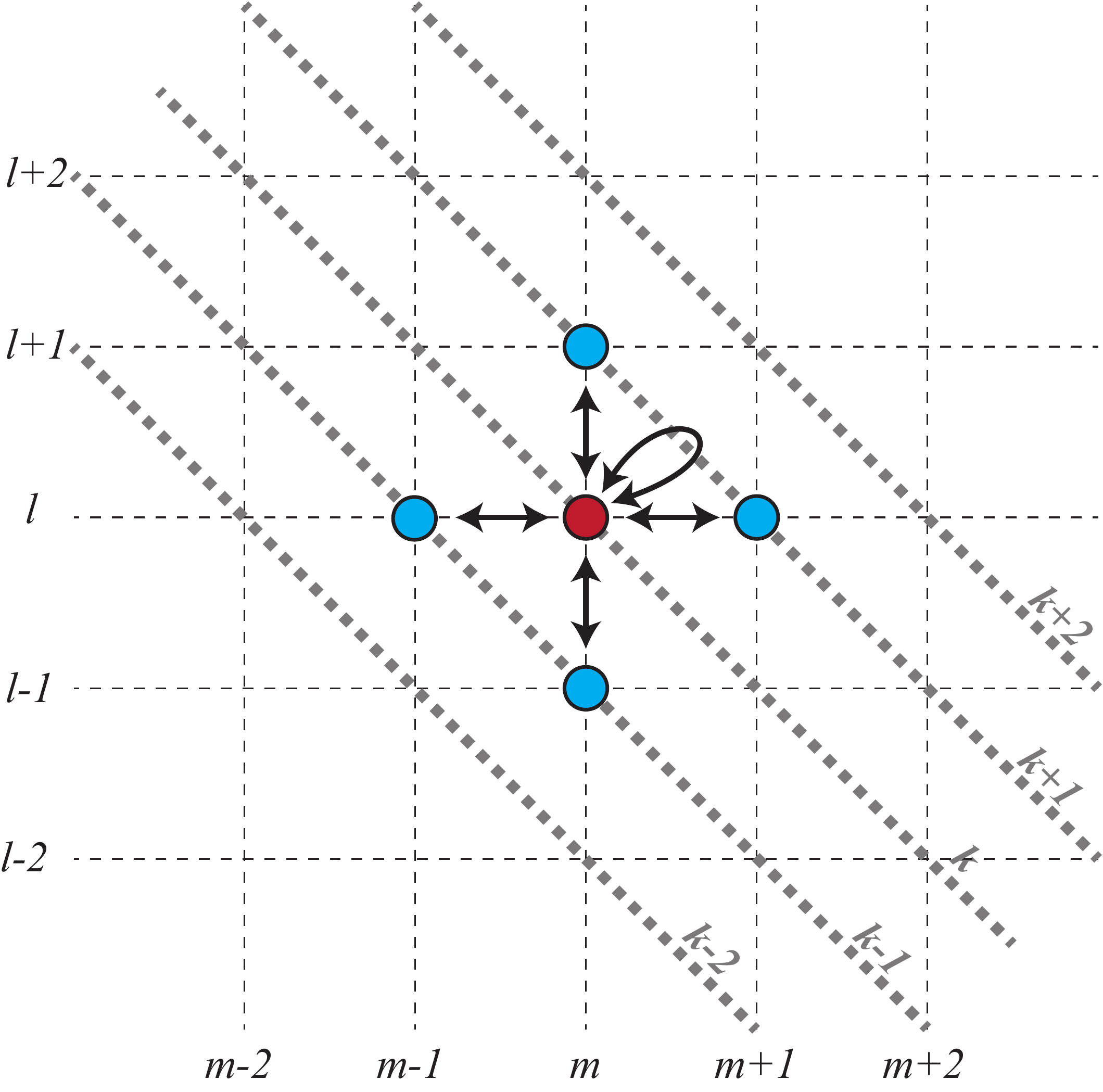}&\includegraphics[width=.5\textwidth]{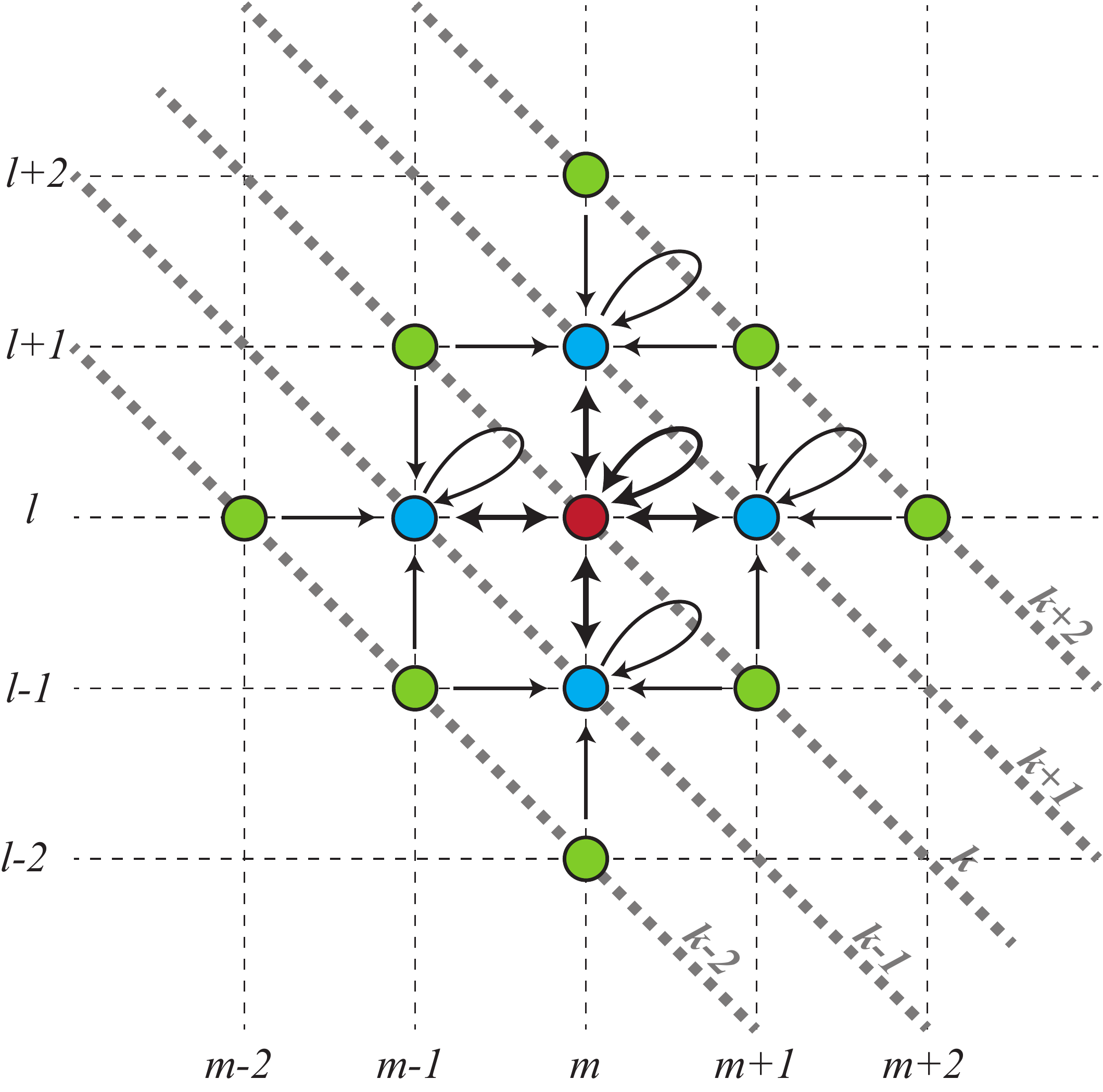}
\end{tabular}
	\caption{Illustration of the possible paths for one $\tilde{X}_{m,l}$ with $m+l=k$ for cylinders of length three (l.h.s) and four (r.h.s). An arrow indicates whether or not one state can be followed by another in a sequence.}
	\label{fig:Meso3Cylinder}
\end{figure}

Let us consider that for the cylinders of length 3.
As noted above, the grammar $\GG_3$ of the system is determined by the fact that $|k_t - k_{t-1}| \leq 1$ and $|k_{t+1}-k_t| \leq 1$. 
Therefore, as illustrated in Fig. \ref{fig:Macro3Cylinder}, for any $k_t = k$ with $0 < k < N$ there are nine possible paths $[k_{t-1},k,k_{t+1}]$ and for $k = 0$ and $k = N$ there are respectively four paths.
In order to compute the probability of a certain macro path, say $[p,k,f]$ $p$ for past and $f$ for future, we have to sum over all meso level paths that contribute to the given macro path.
Let us denote a meso level path as $[(m_p l_p),(m l),(m_f l_f)]$ with $m_p + l_p = p$, $m+l = k$ and $m_f + l_f = f$.
Its probability is given by
\begin{equation}
\mu([(m_p l_p),(m l),(m_f l_f)]) = \tilde{\pi}_{m_p,l_p} \tilde{P}(\tilde{X}_{m_p,l_p},\tilde{X}_{m,l}) \tilde{P}(\tilde{X}_{m,l},\tilde{X}_{m_f,l_f}).
\label{eq:mu.Meso}
\end{equation}
The l.h.s. in Fig. \ref{fig:Meso3Cylinder} illustrates the possible paths for one $\tilde{X}_{m,l}$ with $m+l=k$.
Notice that for a given macro state $X_k$ there are $k+1$ meso states if $k \leq M$ and respectively $N-k + 1$ meso states for $k > M$ (these numbers are for the case $M = L$ with $M+L = N$).
In the case sequences of length three are considered, the situation is still quite clear.
For instance, a macro path $[k-1,k,k+1]$ can be realized in four different ways for each\footnote{Notice that the number of possibilities reduces at the corners or borders of the meso chain whenever $m = 0$ or $l = 0$.} $\tilde{X}_{m,l}$ with $m+l=k$:
\begin{eqnarray}
\label{eq:mesoPaths}
[(m-1 \ l),(m \ l),(m+1 \ l)]\\ \nonumber 
[(m-1 \ l),(m \ l),(m \ l+1)]\\ \nonumber 
[(m \ l-1),(m \ l),(m+1 \ l)]\\ \nonumber  
[(m \ l-1),(m \ l),(m \ l+1)]
\end{eqnarray}
The same reasoning can be applied to derive the probabilities for cylinders of length four even though the situation becomes slightly more complicated, as illustrated on the r.h.s. of Fig. \ref{fig:Meso3Cylinder}.

\begin{figure}[htp]
	\centering
%\begin{tabular}{l r}
\includegraphics[width=.9\textwidth]{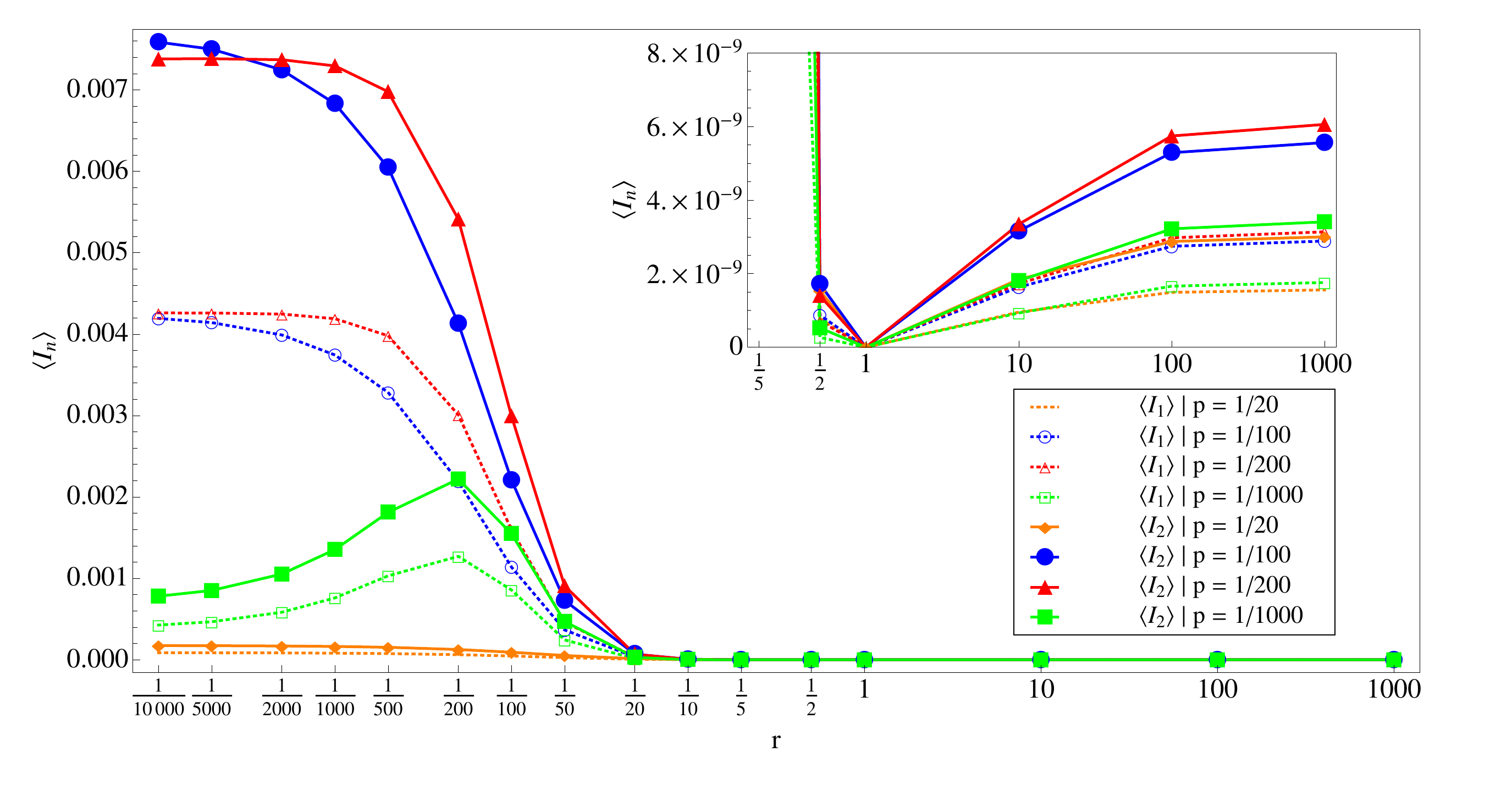}
%\end{tabular}
	\caption{$\langle I_1\rangle$ (dashed curves) and $\langle I_2\rangle$ (solid curves) as a function of the coupling between the two communities $r$ for a system of $N=100$ agents. The different curves represent various different contrarian rates $p$ from $0.05$ to $0.001$, see legend.}
	\label{fig:In.All}
\end{figure}

On the basis of the probabilities of blocks of length three and four respectively, the computation of the Markovianity measures $\langle I_1\rangle = \Delta H_2 - \Delta H_{3}$ and $\langle I_2\rangle = \Delta H_2 - \Delta H_{4}$ is straightforward.
All that is needed is to compute the respective block entropies.
Fig. \ref{fig:In.All} shows $\langle I_1\rangle$ (dashed curves) and $\langle I_2\rangle$ (solid curves) as a function of the coupling between the two communities $r$ for a system of $N=100$ agents ($M = L = 50$).
The different curves represent various different contrarian rates $p$ from $0.001$ to $0.05$.
Notice the log-linear scaling of the figure.

What becomes clear in Fig. \ref{fig:In.All}, first of all, is that the deviation from Markovianity is most significant for small inter-community couplings.
This means, in the reading of \cite{Goernerup2008}, that the information provided by pasts of length $n$ about the future state (beyond that given by the present) is larger than zero for small $r$.
In general and not surprisingly, $\langle I_2\rangle > \langle I_1\rangle$ which means that both the first and the second outcome before the present provide a considerable amount of information.
In fact, the numbers indicate that the first and the second step into the past contribute in almost the same way.
Noteworthy, the two measures $\langle I_1\rangle$ and $\langle I_2\rangle$ behave in the same way from the qualitative point of view which suggests that the computationally less expensive $\langle I_1\rangle$ can be well-suited for the general Markovianity test. %as in \cite{Goernerup2008}.

The inset in Fig. \ref{fig:In.All} shows the situation for values around $r = 1$ (homogeneous mixing) as well as $r > 1$.
As we would expect by the strong lumpability of homogeneous mixing, $\langle I_1\rangle$ and $\langle I_2\rangle$ are zero in the case $r=1$.
Also if the inter-community coupling becomes larger than the coupling within communities (a situation that resembles a bipartite graph) $\langle I_1\rangle$ and $\langle I_2\rangle$ are very small, indicating that a Markovian macro description (i.e., ideal aggregation, cf. \cite{Kemeny1976}, p. 140 and \cite{Buchholz1994}, pp. 61 -- 63) describes well these situations.

\begin{figure}[htp]
	\centering
%\begin{tabular}{l r}
\includegraphics[width=.9\textwidth]{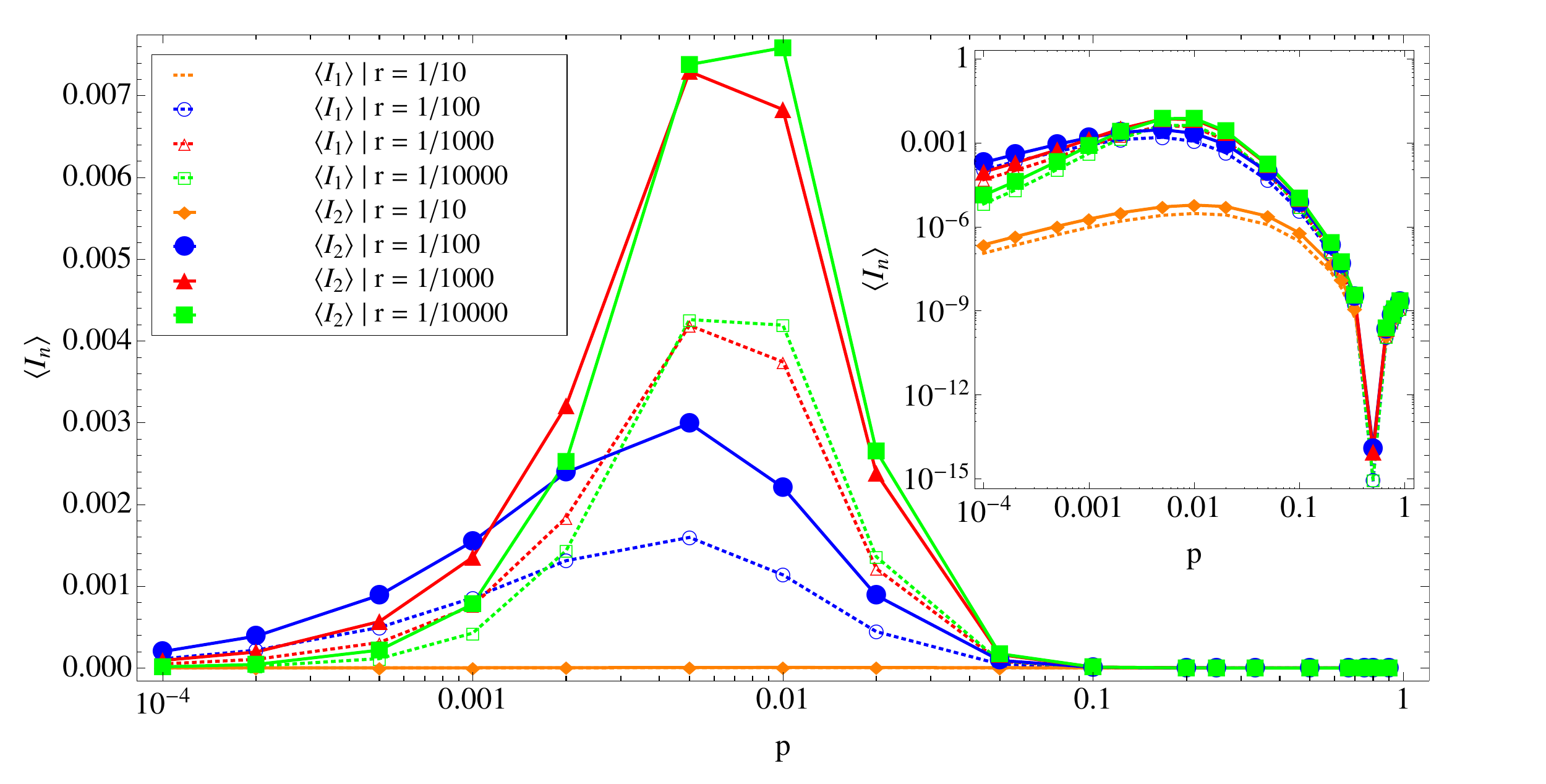}
%\end{tabular}
	\caption{$\langle I_1\rangle$ and $\langle I_2\rangle$ as a function of the contrarian rate $p$ for various coupling ratios $r$ and a system of $M = L = 50$.}
	\label{fig:In.P}
\end{figure}

Finally, we notice in Fig. \ref{fig:In.All} that the measures do not generally increase monotonically with a decreasing ratio $r$ which is most obvious for the example with a very small $p = 1/1000$ (green curves).
This is somewhat unexpected and it indicates the existence of certain parameter constellations at which macroscopic complexity (for this is how non-Markovianity may be read) is maximized. 
To obtain a better understanding of this behavior, the measures $\langle I_1\rangle$ and $\langle I_2\rangle$ are plot in Fig. \ref{fig:In.P} as a function of the contrarian rate $p$.
Notice again the log-linear scaling of the plot.

It becomes clear that there is a strong and non-trivial dependence of the Markovianity measures on the contrarian rate $p$.
Namely, $\langle I_1\rangle$ and $\langle I_2\rangle$ are very small if $p$ is relatively large but they are also relatively small if $p$ becomes very small.
There is a parameter regime in between in which deviations from Markovianity become most significant.
Notice that in the inset of Fig. \ref{fig:In.P} the same curves are shown on a double-logarithmic scale.
This shows, first, that $\langle I_1\rangle$ and $\langle I_2\rangle$ for very small $p$ are still significantly larger compared to the case of relatively large $p$ (say $p > 0.1$). 
Secondly, we observe that $\langle I_1\rangle$ and $\langle I_2\rangle$ actually vanish for $p = 1/2$.
Indeed, it is possibly to show that the two-community CVM with $p = 1/2$ satisfies lumpability conditions  independent of the topological parameter $r$ (cf. \cite{Banisch2014phd},pp.116/17).
%As discussed in the previous section, the reason for that is the strong lumpability of the two-community CVM whenever $p = 1/2$.

\begin{figure}[htp]
	\centering
%\begin{tabular}{l r}
\includegraphics[width=.7\textwidth]{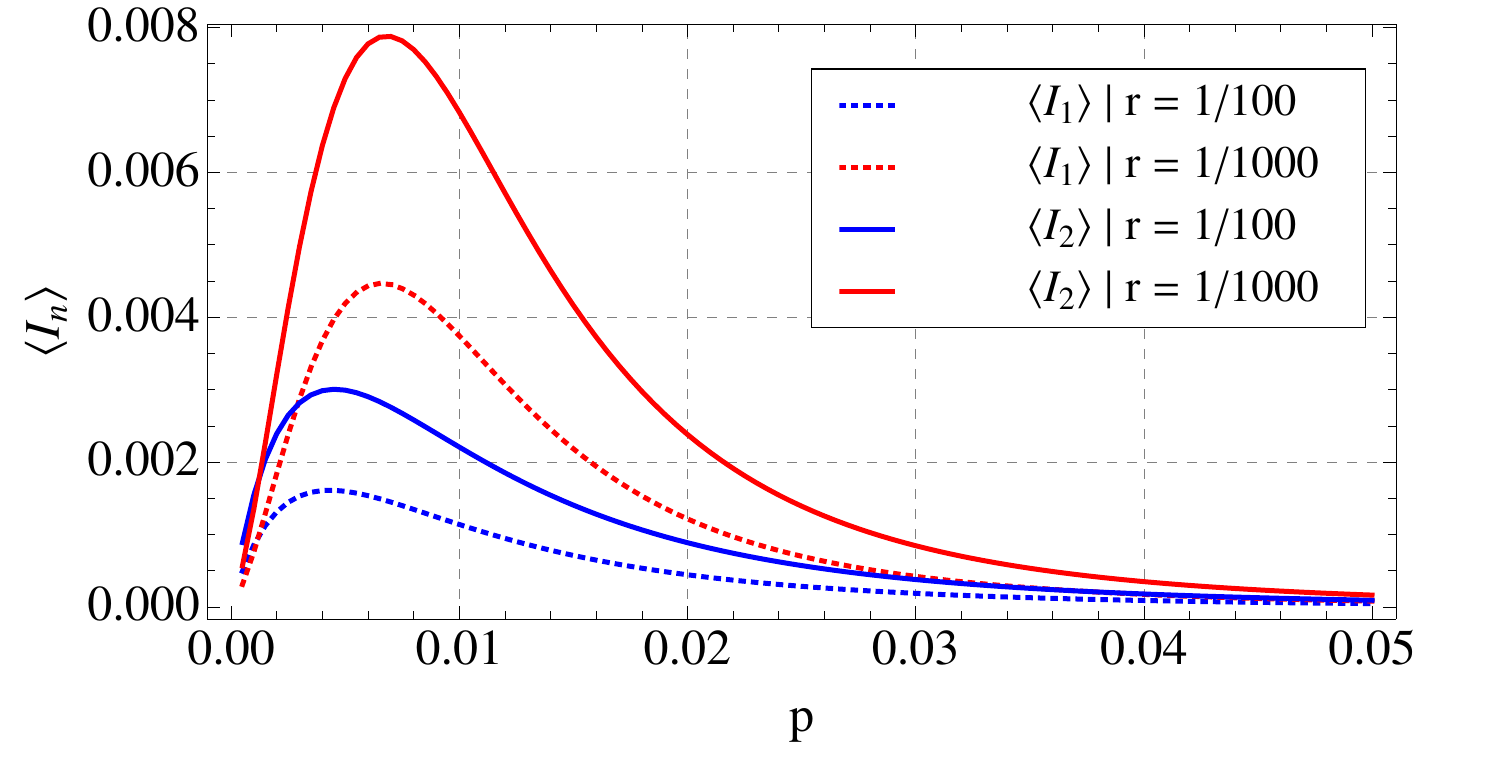}
%\end{tabular}
	\caption{Detailed picture of the dependence of $\langle I_n\rangle$ on the contrarian rate.  Blue curves correspond to $r = 1/100$ and red curves to $r = 1/1000$. In the first case the peak is at around $p \approx 0.05$, in the latter at $p \approx 0.065$.}
	\label{fig:In.P.fine}
\end{figure}

Finally, a detailed picture of the dependence of $\langle I_n\rangle$ on the contrarian rate is provided in Fig. \ref{fig:In.P.fine}.
The plot compares the cases $r = 1/100$ and $r = 1/1000$ in order to show that the peaks in the $\langle I_n\rangle$ depend also on $r$.
For the interpretation of this behavior, notice that the $p$ at which deviations from Markovianity become largest, lie precisely in the parameter interval in which switching times between the two complete consensus states become minimal.
Compare Fig. \ref{fig:CVM.MT.0toN} in Sec. \ref{cha:5.MajMinSwitching}.

\enlargethispage{\baselineskip}
All in all, this analysis shows that global aggregation over an agent population without sensitivity to micro- or mesoscopic structures leads to memory effects at the macroscopic level.

\section{Concluding Remarks}

This paper has provided an analysis of the CVM on the complete and the two-community graph.
Based on the previous work on Markov chain aggregation for ABMs \cite{Banisch2012son,Banisch2013eccs,Banisch2013acs}, higher-level Markov chain descriptions have been derived and allow a detailed understanding of the two cases.
A large contrarian rate $p$ leads to a process which fluctuates around the states with approximately the same number of black and white agents, the fifty-fifty situation $N_{\square} = k = N/2$ being the most probable observation.
This is true for homogeneous mixing as well as for the two-community model.
However, if $p$ is small, a significant difference between the two topologies emerges as the coupling between the two communities becomes weaker.
On the complete graph the population is almost uniform for long periods of time, but due to the random perturbations introduced by the contrarian rule there are rare transitions between the two consensus profiles.
On the community graph, an effect of local alignment is observed in addition to that, because the system is likely to approach a meta-stable state of intra-community consensus but inter-community polarization.

% % REV: link  to related work

A order-disorder phase transition as the contrarian rate increases has been observed on the complete graph in several previous contrarian opinion models (e.g., \cite{Galam2004,Lama2005,Sznajd-Weron2011,Nyczka2012}).
For the CVM, in the transition from consensus switching to disorder there is a phase in which the process leads uniform stationary distribution in which all opinion frequency levels $0 \leq k \leq N$ are observed with equal probability ($\pi_k = 1/(N+1)$).
The contrarian rate $p$ at which this happens is $p^* = 1/(N+1)$ and depends inversely on the system size such that a model with a single contrarian agent fails to enter the ordered regime.
This confirms and explains the behavior observed in \cite{Masuda2013} for a model with a fixed number of contrarian agents.

A particular focus of this paper has been on the effect of inhomogeneities in the interaction topology on the stationary behavior.
In this regard, the two-community CVM served as a suitable scenario to assess the macroscopic effects introduced by a slight microscopic heterogeneity.
Namely, homogeneous mixing compatible with the usual way of aggregation over all agents leads to a random walk on the line with $N+1 = O(N)$ states whereas the two-community model leads to a random walk on a 2D lattice with $O(N^2)$ states.
As the latter is a proper refinement of the former this gives us means to study the relation between the two coarse-grainings in a Markov chain setting.
In this regard, this paper has made visible the reasons for which lumpability fails, %even in its weaker form (Sec. \ref{cha:5.WeakLumpCVM}).
and it has also provided a first analysis of the macroscopic memory effects that are introduced by heterogeneous interaction structures.
In this regard, the paper demonstrates that information-theoretic measures are a promising tool to study the relationship between different levels of description in ABMs.

There are various issues that deserve further discussion.
For instance, is the emergence of memory in the transition from the micro to the macro level a useful characterization for the complexity of a certain system?
Namely, the theory of Markov chain aggregation makes explicit statements about when a micro process is \emph{compressible} to a certain macro-level description.
This links non-lumpability to computational incompressibility, one of the key concepts in dynamical emergence \cite[among others]{Bedau2003,Huneman2008}.
This point shall be discussed in a forthcoming paper.
%We shall discuss this point in the next chapter.

Finally, I would like to mention the possibility of applying the arguments developed in the second part this paper to the case of models with absorbing states as, for instance, the pure VM ($p=0$).
In that case, the quasi-stationary distribution (see \cite{Darroch1965}) takes the role of $\hat{\pi}$ or respectively $\tilde{\pi}$ in %the construction of an ideal aggregate and 
the computation of cylinder measures.
One interesting issue to be addressed in this regard is to reconsider the question of weak lumpability for the VM. %because it is well-known that the long-term dynamics of the VM on the ring is a random walk.
Finally, to understand how microscopic heterogeneity and macroscopic complexity are related, numerical experiments with different network topologies are another promising way to continue the analysis started in this paper.

%\bibliographystyle{ws-acs}
%\bibliography{../../REFs/SynThese.bib}

\end{document}